\documentclass[
aps,prd,
reprint,
superscriptaddress,
nofootinbib,
nobibnotes,
amsmath,amssymb,
]{revtex4-2}

\usepackage{aas_macros}
\usepackage{graphicx}
\usepackage{dcolumn}
\usepackage{bm}
\usepackage{hyperref}
\usepackage{mathtools}
\usepackage{xcolor}
\usepackage{multirow}
\usepackage{enumitem}

\newcommand\T{\rule{0pt}{2.6ex}} 
\newcommand\B{\rule[-1.2ex]{0pt}{0pt}} 

\newcommand{\LCDM}{$\rm\Lambda CDM$}
\newcommand{\LeDM}{$\rm\Lambda eDM$}
\newcommand{\sig}{$\sigma_{8}$}
\newcommand{\ie}{\emph{i.e.}}
\newcommand{\anz}{a_{\rm nz}}

\begin{document}

\preprint{APS/123-QED}

\title{Dark matter solution to the $H_{0}$ and $S_{8}$ tensions,\\ and the integrated Sachs-Wolfe void anomaly}

\author{Krishna Naidoo}
\email[]{k.naidoo@ucl.ac.uk}
\affiliation{Center for Theoretical Physics, Polish Academy of Sciences, al.\,Lotnik\'{o}w 32/46 Warsaw, Poland}
\affiliation{Department of Physics \& Astronomy, University College London, Gower Street, London, WC1E 6BT, UK}
\author{Mariana Jaber}
\affiliation{Center for Theoretical Physics, Polish Academy of Sciences, al.\,Lotnik\'{o}w 32/46 Warsaw, Poland}
\author{Wojciech A. Hellwing}
\affiliation{Center for Theoretical Physics, Polish Academy of Sciences, al.\,Lotnik\'{o}w 32/46 Warsaw, Poland}
\author{Maciej Bilicki}
\affiliation{Center for Theoretical Physics, Polish Academy of Sciences, al.\,Lotnik\'{o}w 32/46 Warsaw, Poland}
\date{\today}

\begin{abstract}
	We consider a phenomenological model of dark matter with an equation-of-state $w$ that is negative and changing at late times. We show this couples the $H_{0}$ and $S_{8}$ tensions, alleviating them both simultaneously, reducing the $H_{0}$ tension from $\sim5\sigma$ to $\sim3\sigma$ and the $S_{8}$ tension from $\sim3\sigma$ to $\sim1\sigma$. Furthermore, the model provides an explanation for the anomalously large integrated Sachs-Wolfe (ISW) effect from cosmic voids, a unique consequence of the changing and negative equation-of-state. Observations of high ISW from cosmic voids may therefore be evidence that dark matter plays a significant role in both the $H_{0}$ and $S_{8}$ tensions. We predict the ISW from cosmic voids to be a factor of up to $\sim2$ greater in this model than what is expected from the standard model \LCDM{}. These results extend to other degenerate models of dark matter, such as unified or interacting dark matter and dark energy models.
\end{abstract}

\maketitle


\section{Introduction}

The emergence of strong tensions in the constraints of the Hubble constant $H_{0}$ and $S_{8}$ parameters, between early \citep{Planck2018} and late universe-based observations \citep{Riess2016, Kids2021Asgari, DES2022, DESKids2023, HSC2023}, have placed intense scrutiny on observations, methods and the assumptions of the standard model of cosmology \LCDM{} (see \cite{Verde2019, Knox2020, Eleonora2021, Jedamzik2021, Eleonorab2021, Perivolaropoulos2022, Shah2021, Abdalla2022} for reviews on the Hubble tension). One of these assumptions is the presence of cold dark matter; a yet-to-be-detected massive particle that is the dominant source of gravitation in the universe. Dark matter's interactions with standard model particles and forces is known to be very weak and the specific properties of dark matter are often assumed to manifest only on small cosmological scales.

However, with no direct observations of dark matter, there is little we can presume about its properties, other than its gravitational effects and weak interactions with known particles. Rather than consider an endless list of dark matter model extensions we can instead consider phenomenological scenarios that allow us to determine the general observational implications of a whole family of models, including their role in tensions and anomalies. With this in mind, we explore the implication of a subset of the generalised dark matter model \citep{Hu1998}, in a spatially flat universe with a cosmological constant ($\Lambda$). The equation-of-state (EoS) for dark matter ($w_{\rm dm}$) is allowed to be non-zero and evolving at late times but with null speed of sound and viscosity. For consistency with constraints from the early universe \citep{Xu2013, Kopp2018} the EoS is assumed to be effectively zero at early times. We will refer to this model as \emph{evolving} dark matter (eDM), to distinguish it from other models often abbreviated to WDM (such as warm dark matter), and with the full model with $\Lambda$ referred to as \LeDM{}.

While the $H_{0}$ and $S_{8}$ tensions are widely discussed in the community, a lesser known anomaly is the observation of larger than expected integrated Sachs-Wolfe \citep[ISW;][]{Sachswolfe1967} signal from cosmic voids \citep{Granett2008, Cai2014, Kovacs2018} (the void-ISW anomaly). This anomaly is strongest for `photometric' voids, \ie{} voids measured from photometric observations of galaxies which are preferentially elongated along the line-of-sight (LOS) \citep{Kovacs2019} and ranging in significance from 2-4$\sigma$ \cite{Granett2008, Cai2014, Kovacs2019}, while for `spectroscopic' voids (smaller and not aligned with the LOS) the ISW is larger but to a lesser extent and remains consistent with \LCDM{} \citep{specz_params3}. In contrast, observations of void lensing are either consistent with \LCDM{} or lower than expected \citep{Kovacs2022}.

The aim of this paper is to establish the role dark matter can play in alleviating tensions in cosmology and in explaining the void-ISW anomaly. The latter has rarely been considered part of the tension discussion, but in this paper we show the ISW is very sensitive to late-time changes in dark matter's EoS and can provide a unique test for a dark matter (or an interacting dark sector) solution to tensions. This paper is organised as follows: (1) we discuss the assumptions and theoretical background of \LeDM{} and the implications for the $H_{0}$ and $S_{8}$ tensions, and the ISW and lensing from cosmic voids; (2) we describe \LeDM{}'s parameter degeneracies and methods used for parameter inference, (3) we discuss the constraints on \LeDM{} parameters using cosmological observations; and (4) we discuss the implications for tensions and anomalies in cosmology.


\section{Theory}

\subsection{\LeDM{} and Tensions}

The evolution of the dark matter density $\rho_{\rm dm}$ can be described as a function of the scale factor $a$ and a time-varying EoS $w_{\rm dm}(a)$ defined as
\begin{align}
	& \rho_{\rm dm}(a) = \frac{\rho_{\rm dm,0}}{a^{3}}\,W(a),\\
	& W(a) = \exp\left(3\int_{a}^{1}\frac{w_{\rm dm}(a')}{a'} {\rm d}a'\right),
\end{align}
where $\rho_{\rm dm,0}$ is the dark matter density today and $a'$ a dummy variable for the integration by the scale factor. This general solution is derived from the assumption of energy conservation and the continuity equation. We consider the following function for the EoS,
\begin{equation}
	w_{\rm dm}(a)=
	\begin{dcases}
		w_{\rm dm,0}\left(\frac{a-\anz}{1-\anz}\right),\,\,&\text{for }a\geq\anz,\\
		0 \vphantom{\frac{0}{0}},&\text{otherwise,}\\
	\end{dcases}
\end{equation}
\ie{} a linearly increasing/decreasing EoS where $w_{\rm dm,0}$ is the current value of the dark matter EoS and $\anz$ the scale factor at which dark matter's EoS becomes non-zero.  At early times (i.e. for $a < \anz$) constraints from CMB measurements have shown the EoS must be very close to zero ($|w_{\rm dm}| \lesssim 0.001$) \cite{Xu2013,Kopp2018}, however, at late times the constraints are more relaxed \cite{Kopp2018}. For this reason we fix the EoS at early times to zero and allow the EoS to be non-zero at late times. However, this means the EoS is non-differentiable at $\anz$, a detail that should be revisited in future studies of the model. Generalised dark matter models with a cosmological constant, like the model considered here, are equivalent to interacting dark energy models \cite{Cesare2022}, models where dark matter interacts with dark energy. We can therefore think of this model describing a dark matter-dark energy interaction \cite{Farrar2004}, which could occur only at low energies or densities, or through the decay of unstable dark matter particles with very long half-lives (see \cite{Peebles2012} for the evanescent matter model with similar properties). In both cases the interaction is only relevant at very late-times and would grow over time, following the behaviour of \LeDM{}. In section~\ref{sec_model_degeneracies} we will show the sensitivity to $w_{\rm dm,0}$ is much stronger than $\anz$ and that the parameters are strongly degenerate. We break this degeneracy by setting $\anz=0.5$, meaning dark matter's EoS only becomes non-zero at $z\leq1$.

\begin{figure*}
	\includegraphics[width=\textwidth]{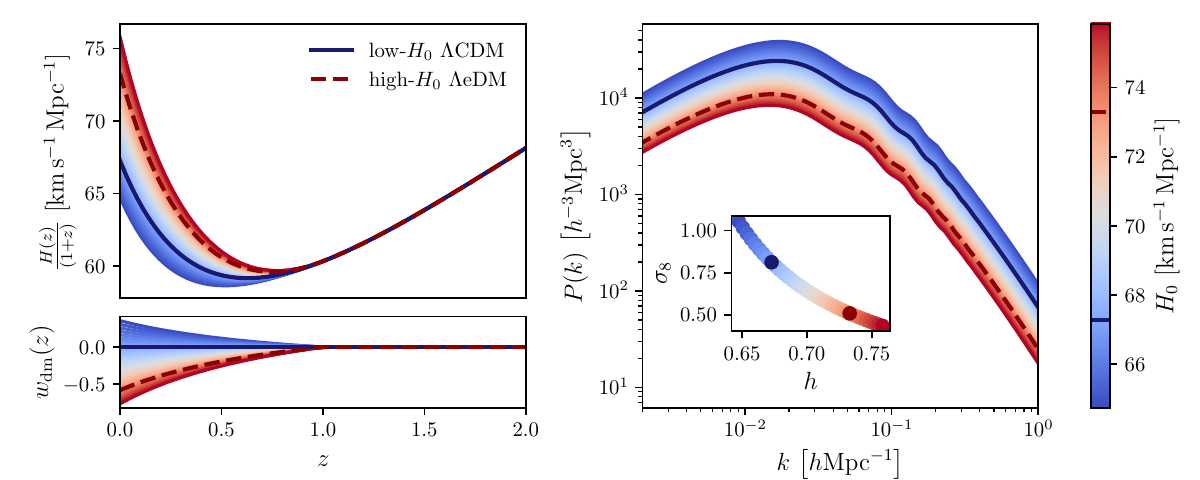}
	\caption{The Hubble function $H(z)$, dark matter EoS $w_{\rm dm}(z)$ and power spectrum $P(k)$ are displayed for models of \LeDM{} as a function of the Hubble constant $H_{0}$, with early universe physics fixed to CMB constraints. A higher $H_{0}$ requires a negative EoS for dark matter, which leads to a decrease in the $P(k)$ amplitude and therefore a low \sig{} (shown in the inset plot on the right). Therefore, \LeDM{} can alleviate both the $H_{0}$ and $S_{8}$ tensions as a high-$H_{0}$ is naturally coupled to a low-\sig{}. For reference the profiles for \LCDM{} assuming early universe CMB constraints are shown with dark blue solid lines and the profiles for \LeDM{} assuming late universe constraints on $H_{0}$ are shown with red dashed lines. \label{fig_pk_hz}}
\end{figure*}

The total matter density is defined by
\begin{equation}
	\rho_{\rm m}(a) = \frac{\rho_{\rm m, 0}}{a^{3}}\,\Gamma(a),
\end{equation}
with the additional term
\begin{equation}
	\Gamma(a) = f_{\rm b} + f_{\rm \nu} + f_{\rm dm}\,W(a).
\end{equation}
Here $f_{\rm i}$ is the fraction of the matter constituent $\rm i$ to the total matter content today (\ie{}~$f_{\rm i}=\rho_{\rm i,0}/\rho_{\rm m, 0}$); the subscripts $\rm b$ indicate baryons, $\nu$ non-relativistic neutrinos and $\rm dm$ dark matter. Another useful relation is the derivative of $\Gamma(a)$ with respect to $\ln a$, which is given by
\begin{equation}
	\Upsilon(a) = \frac{{\rm d}\,\Gamma(a)}{{\rm d}\ln a} = 3\, f_{\rm dm}\,\frac{w_{\rm dm}(a)\, W(a)}{\Gamma(a)}.
\end{equation}
In \LCDM{} these functions simplify to $\Gamma(a)=1$ and $\Upsilon(a)=0$.

We modify the background evolution and the linear perturbation equations of dark matter in the cosmological Boltzmann solver \texttt{CLASS} \citep{CLASS}, following the above relations and equation 2 of \citep{Kopp2018}. From \texttt{CLASS} we output the power spectrum and growth functions.

To understand the role that \LeDM{} may play in the $H_{0}$ tension we fix the conditions of the early universe to those inferred by Planck CMB measurements \citep{Planck2018}. This is carried out by fixing the primordial power spectrum, optical depth and the physical densities of baryons and dark matter to $A_{\rm s}=2.101\times10^{-9}$, $n_{\rm s}=0.9649$, $\tau=0.0544$ and $\Omega_{\rm b}h^{2}=0.02236$, respectively. For dark matter this means fixing its value at early times to
\begin{equation}
	\Omega_{\rm dm}W_{0}h^{2} = \Omega_{\rm dm}^{\rm init}h^{2} = 0.1202,
\end{equation}
where $\Omega_{\rm dm}^{\rm init}$ is the fractional density of dark matter if the EoS is zero and $W_{0}=W(\anz)$.

In Fig.~\ref{fig_pk_hz} we fix the parameters of \LeDM{} as described above while varying a single parameter -- the Hubble constant $H_{0}$. In this plot we show that changing $H_{0}$ to values above CMB constraints (of $H_{0}=67.27\,{\rm km\, s^{-1}\, Mpc^{-1}}$ indicated with a solid blue line) requires dark matter's EoS to be negative. A consequence of a high $H_{0}$ is a decrease in the power spectrum amplitude and therefore low \sig{}. Since $\sigma_{8}\propto S_{8}$ this naturally couples the two tensions. If $H_{0}$ is set to the values observed by SH0ES \citep{Riess2021} (of $H_{0}=73.3\,{\rm km\, s^{-1}\, Mpc^{-1}}$) we obtain \sig{}$\sim0.5$.

\subsection{Implications for the ISW and lensing}

The ISW $T_{\rm ISW}$ is defined as
\begin{equation}
	\frac{T_{\rm ISW}(\hat{\eta})}{T}=\frac{2}{c^{3}}\int_{0}^{\chi_{\rm LS}} \dot{\Phi}\left[\chi\hat{\eta},t(\chi)\right]\,a(\chi)\,{\rm d}\chi,
	\label{eq_ISW}
\end{equation}
where $\dot{\Phi}$ is the time-derivative of the gravitational potential $\Phi$, $c$ is the speed of light, $\chi$ is the transverse comoving distance in a direction $\hat{\eta}$ and $\chi_{\rm LS}$ is the comoving distance to the last scattering surface. The gravitational potential is related to the density contrast $\delta$ by the Poisson equation which in the linear regime simplifies to
\begin{equation}
	\Phi(x, t)=\frac{3}{2}H_{0}^{2}\,\Omega_{\rm m}\,\Gamma(t)\,\frac{D(t)}{a(t)}\,\nabla^{-2}\delta(x,0),
	\label{eq_phi}
\end{equation}
where $\Omega_{\rm m}$ is the total fractional matter density today, $D(t)$ the linear growth function and $\delta(x)$ the density contrast today (\ie{}~at $z=0$). Consequently, the time-derivative is given by
\begin{equation}
	\dot{\Phi}(x,t)=H(t)\, \Big[f(t)-1+\Upsilon(t)\Big]\Phi(x,t).
	\label{eq_phidot}
\end{equation}
where $f(t)$ is the linear growth rate.

The lensing convergence is given by
\begin{equation}
	\kappa(\hat{\eta}) = \frac{3H_{0}^{2}}{2c^{2}}\,\Omega_{\rm m}\,\int_{0}^{\chi_{\rm s}} \frac{\chi}{\chi_{\rm s}}\,(\chi_{\rm s}-\chi)\frac{\Gamma(\chi)}{a(\chi)}\,\delta(\chi)\,{\rm d}\chi,
	\label{eq_kappa}
\end{equation}
where $\chi_{\rm s}$ is the comoving distance to the lensing source.

Both the ISW and lensing are dependent on $\Gamma(t)$ (see Eqs.~\ref{eq_ISW},  \ref{eq_phi},  \ref{eq_phidot} and \ref{eq_kappa}), while the ISW is also dependent on $\Upsilon(t)$ (see Eq.~\ref{eq_ISW} and Eq.~\ref{eq_phidot}). This means the ISW and lensing signals will be different in \LeDM{} than in \LCDM{} and can be constrained from studies cross-correlating galaxy surveys with weak lensing and CMB temperature maps. However, cross-correlation studies of the ISW typically have very large errors since the signal is highest on large angular scales and subject to cosmic variance (see \cite{Hang2021}). This makes distinguishing the effect of \LeDM{} this way rather challenging. Voids on the other hand allow us to improve the signal-to-noise by stacking void-ISW signals. They are extremely sensitive to \LeDM{} because they accumulate the departures in $\Gamma(t)$ and $\Upsilon(t)$ from \LCDM{}; an effect which is more relevant for larger voids elongated along the LOS, such as those measured by photometric surveys.

The ISW and $\kappa$ profiles for three types of voids are explored, assuming the elliptical void profile of \citep{EllipticalVoid}: (1) spec-$z$ voids refers to typical voids found from spectroscopic surveys \citep{specz_params1, specz_params2, specz_params3, specz_params4}; (2) photo-$z$ voids from photometric surveys \cite{photoz_params1, Kovacs2019, Kovacs2022}, \ie{} large and preferentially elongated along the LOS; and (3) the combined profiles of the Eridanus voids \citep{Mackenzie2017} which are located in the direction of the CMB Cold Spot anomaly \citep{Vielva2004} and for which the role of the ISW from voids has generated considerable discussion \citep{Inoue2006, Inoue2007, Masina2009, Nadathur2014, EllipticalVoid, Naidoo2016, Caballero2016, Naidoo2017, Kovacs2018, Kovacs2020, KovacsCS2022}. The void parameters used and the measured ISW and lensing amplitude with respect to \LCDM{} are given in Table~\ref{tab_voids}.

\begin{table}
	\caption{We list the void parameters for spectroscopic voids (spec-$z$) \cite{specz_params1, specz_params2, specz_params3, specz_params4}, photometric voids (photo-$z$) \cite{photoz_params1, Kovacs2019, Kovacs2022} and the Eridanus voids (combining profiles from voids E-1, E-2, E-3 and E-4) \cite{Mackenzie2017}. The void parameters are the central redshift $z_{\rm c}$, the central density contrast $\delta_{0}$, the effective radius $R$ and the measured amplitude of the ISW $A_{\rm ISW}$ and lensing $A_{\kappa}$ signals with respect to \LCDM{}. Note the measured amplitudes are omitted for the Eridanus voids as this corresponds to a singular location on the sky. Furthermore all voids are assumed to be spherical, with the exception of the photo-$z$ voids which are preferentially aligned towards the LOS with a LOS and perpendicular ratio of $R_{\parallel}/R_{\perp} = 2.6$ \citep{photoz_params1}.\label{tab_voids}}
	\begin{center}
		\begin{tabular}{lcccccc}
			\hline
			\T \B Void\quad\quad & $\quad\delta_{0}\quad$ & $\quad z_{\rm c}\quad$ & $R$ $[h^{-1}{\rm Mpc}]$ & $A_{\rm ISW}$ & $A_{\kappa}$\\
			\hline
			\T Spec-$z$  & $-0.6$ & $0.6$ & $35$ & $1.64\pm0.53$ & $0.97\pm0.19$\\[-5pt]
			\multicolumn{6}{l}{\dotfill}\\
			Photo-$z$ & $-0.55$ & $0.5$ & $60$ & $4.1\pm2$ & $0.79\pm0.12$\\[-5pt]
			\multicolumn{6}{l}{\dotfill}\\
			E-1 & $-0.34$ & $0.14$ & $119$ & n/a & n/a \\
			E-2 & $-0.87$ & $0.26$ & $50$ & n/a & n/a \\
			E-3 & $-0.8$ & $0.3$  & $59$ & n/a & n/a \\
			\B E-4 & $-0.62$ & $0.42$ & $168$ & n/a & n/a \\
			\hline
		\end{tabular}
	\end{center}
\end{table}

\begin{figure*}
	\includegraphics[width=\textwidth]{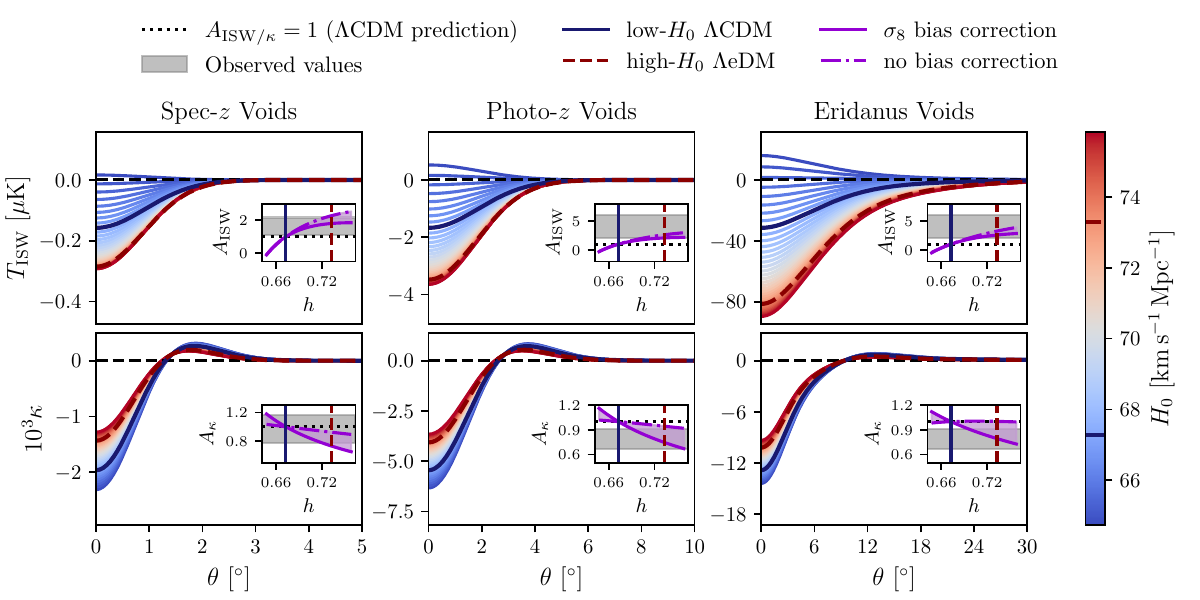}
	\caption{The ISW (top panels) and $\kappa$ (bottom panels) angular profiles for three void types (spec-$z$ on the left, photo-$z$ in the middle and Eridanus voids on the right) are shown as a function of $H_{0}$ in \LeDM{} with all other parameters fixed to early universe CMB constraints. In the inset plots we show the amplitude of the ISW and $\kappa$ profiles with respect to \LCDM{}, shown with a \sig{} bias correction (solid purple lines) and without a bias correction (dashed purple lines). Measured values are indicated with horizontal grey bands. A higher $H_{0}$ resolves the void-ISW anomaly and recovers lower $\kappa$ amplitudes. For reference the profiles for \LCDM{} assuming early universe CMB constraints are shown with dark blue solid lines and the profiles for \LeDM{} assuming late universe constraints on $H_{0}$ are shown with red dashed lines.\label{fig_isw_kappa}}
\end{figure*}

Measurements of the void density profiles are based on observations of galaxies, biased tracers of the underlying density field. To correct for the galaxy bias, a linear galaxy bias parameter is usually fitted based on clustering analysis. However, if \LCDM{} is not the correct model, the inferred bias will be incorrect. By making the approximation that the $P(k)$ for \LeDM{} can be related to the $P(k)$ in \LCDM{} by an amplitude shift, we can make a \sig{} bias correction to the central void density $\delta_{0}$ (the central density contrast obtained from \LCDM{}) by multiplying by the factor $\sqrt{\sigma_{8}^{\Lambda {\rm CDM}}/\sigma_{8}^{\Lambda {\rm eDM}}}$ where $\sigma_{8}^{\Lambda {\rm CDM}}$ is the \sig{} obtained from \LCDM{} and $\sigma_{8}^{\Lambda {\rm eDM}}$ from \LeDM{}.

In Fig.~\ref{fig_isw_kappa} we show the void ISW and lensing profiles as a function of $H_{0}$. As was shown in Fig~\ref{fig_pk_hz}, an increase in $H_{0}$ requires a negative and decreasing EoS for dark matter, resulting in $\Gamma\neq1$ and $\Upsilon\neq0$. This results in a significant increase in the ISW profile and a slight decrease in lensing, fitting well with observational measurements (indicated in the inset plots). For spec-$z$ and photo-$z$ voids the ISW and lensing measurements were made by stacking voids, something which cannot be performed for the Eridanus voids which lies on a singular patch of sky. For this reason and because these voids more closely resemble photo-$z$ voids we will assume that their ISW and lensing profiles follow the relation for photo-$z$ voids.

\section{Methods}

\subsection{Model Degeneracies\label{sec_model_degeneracies}}

\begin{figure*}
	\includegraphics[width=0.99\columnwidth]{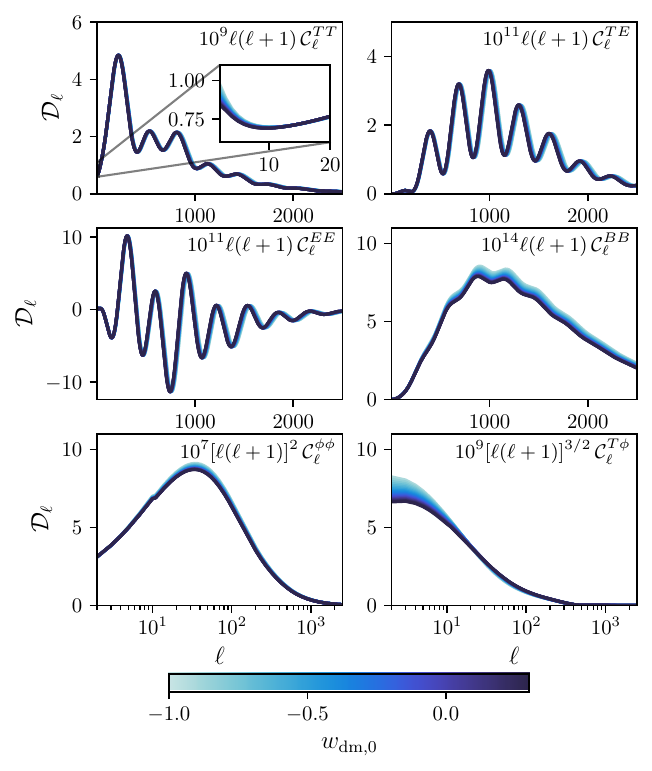}
	\includegraphics[width=0.99\columnwidth]{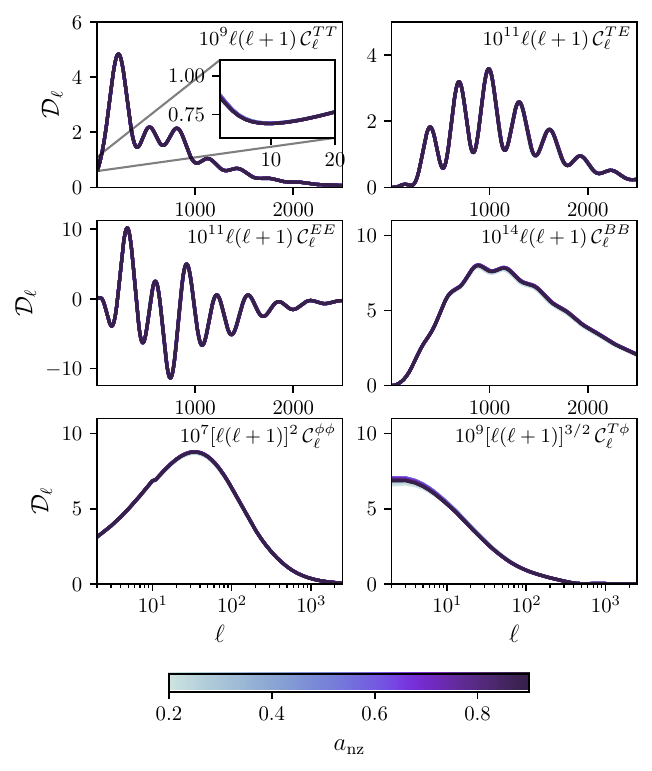}
	\caption{The angular power spectra $\mathcal{D}_{\ell}$ are shown as a function of $w_{\rm dm,0}$ on the left and $\anz$ on the right, with all other parameters held fixed to P18 \LCDM{} cosmology and $\anz=0.5$. In the subplots we show the CMB temperature autocorrelation ($TT$, top left), temperature to E-mode cross correlation ($TE$, top right), E-mode autocorrelation ($EE$, middle left), B-mode autocorrelation ($BB$, middle right), lensing potential autocorrelation ($\phi\phi$, bottom left) and temperature to lensing potential cross correlation ($T\phi$, bottom right). The text in each subplot shows the multiplication factor used for plotting each angular power spectra. \label{fig_w_dm_a_nz}}
\end{figure*}

\begin{figure*}
	\includegraphics[width=\textwidth]{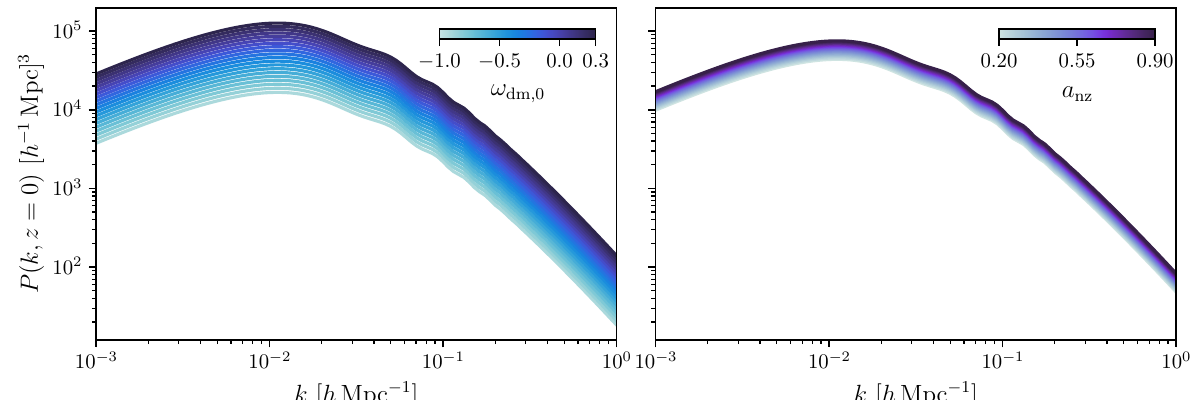}
	\caption{The power spectrum at redshift zero for \LeDM{}, with varying $w_{\rm dm,0}$ and $\anz=0.5$ and on the left and varying $\anz$ with $w_{\rm dm,0}=-0.2$ on the right. The plots show a strong degeneracy between these parameters, smaller values of $\anz$ and $w_{\rm dm,0}$ dampen $P(k)$ more strongly, an effect which is more sensitive to $w_{\rm dm,0}$. \label{fig_pk_wdm_anz}}
\end{figure*}

The \LeDM{} model is defined with the addition of two free parameters, $w_{\rm dm,0}$ which defines the dark matter EoS today (i.e.~at $z=0$) and $\anz$ the scale factor at which the EoS becomes non-zero. In Fig.~\ref{fig_w_dm_a_nz} we show the angular power spectra of the CMB temperature, E and B-modes, and lensing potentials for the \LeDM{} model with base parameters (i.e.~those in common with \LCDM{}) fixed to best-fit \LCDM{} P18 constraints. In Fig.~\ref{fig_pk_wdm_anz} we show the power spectra at redshift $z=0$ with base parameters fixed to best-fit \LCDM{}. In Fig.~\ref{fig_w_dm_a_nz} and \ref{fig_pk_wdm_anz} the analytical angular and 3D power spectra are shown as a function of $w_{\rm dm,0}$ varying between $-1$ and $0.3$ with $\anz=0.5$ and separately as a function of $\anz$ varying between $0.2$ and $0.9$ with $w_{\rm dm,0}=-0.2$.

We see that smaller $w_{\rm dm,0}$ and $\anz$ lead to a very similar dampening effect on the power spectra, a shift to the right in the temperature and E-mode angular auto/cross correlation, and an amplification to the B-mode and lensing auto/cross-correlations. At low-$\ell$ we see a modest increase to the temperature autocorrelation (highlighted in the inset plots), this is caused by larger ISW in these models. These figures illustrate the strong degeneracy between these two parameters and that cosmological probes will be much more sensitive to $w_{\rm dm,0}$ than $\anz$. For this reason we fix $\anz$ to $0.5$ in our analysis.

\subsection{Parameter Inference}

Constraints on cosmological parameters were obtained using the cosmological Markov chain Monte Carlo (MCMC) software \texttt{Cobaya} \cite{CobayaSoftware2019, CobayaPaper2021}. This was done using a combination of Planck CMB measurements, Pantheon Type Ia supernovae, constraints on the absolute magnitude of Type Ia supernovae, baryonic acoustic oscillations and redshift space distortion measurements. Specifically:
\begin{enumerate}[label=(\roman*)]
	\setlength\itemsep{-0.25em}
	\item Planck CMB measurements of temperature, polarisation auto/cross correlation for high-$\ell$ (i.e.~$\ell>30$), temperature and polarisation auto correlation for low-$\ell$ (i.e.~$\ell \leq 30$) and lensing \cite{Aghanim2019}. The joint likelihood for Planck CMB measurements are referred to in the paper as P18.
	\item Pantheon Type Ia supernovae \cite{Scolnic2017} referred to as SN.
	\item SH0ES Cepheid constraints on the Type Ia absolute magnitude \cite{RiessMb2021}, referred to as SN+$M_{B}$.
	\item Baryonic acoustic oscillations and redshift space distortion measurements from 6dF, SDSS DR7 and BOSS \citep{Beutler2012, Ross2014, Alam2016}, which are referred to as BAO.
\end{enumerate}

\begin{figure*}
	\includegraphics[width=\textwidth]{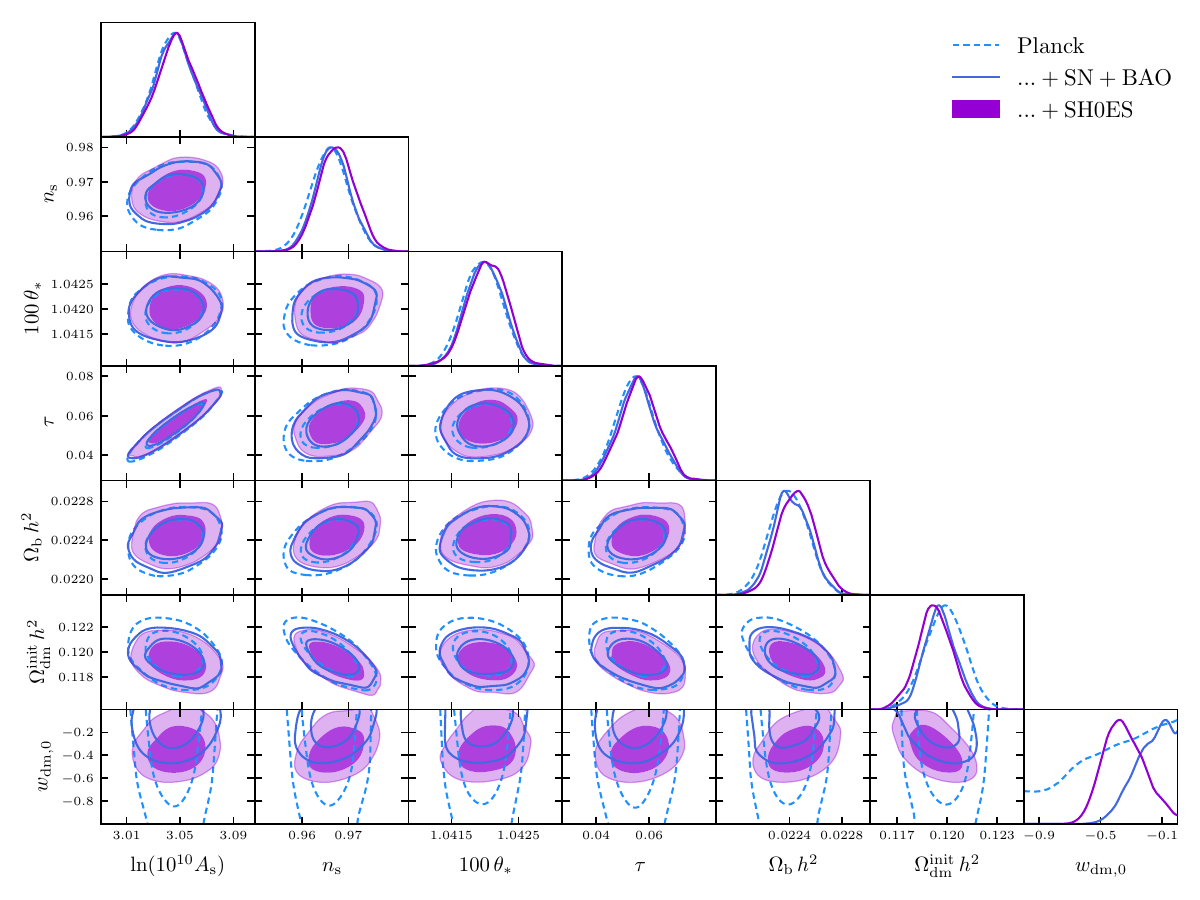}
	\caption{Constraints on \LeDM{} parameters for P18 (dashed blue lines), P18+SN+$M_{B}$ (solid blue lines) and P18+SN+$M_{B}$+BAO (filled purple). The contours represent the 68\% and 95\% confidence regions. P18 does not strongly constrain $w_{\rm dm,0}$, but the addition of SN+$M_{B}$ and BAO pulls it away from $-1$ and towards $-0.3$, although these constraints remain consistent with zero.\label{fig_params}}
\end{figure*}

The constraints for parameters from \LCDM{} and \LeDM{} are obtained by sampling the posterior by running an MCMC. In both cases, this means varying the standard \LCDM{} parameters, i.e.~primordial power spectrum amplitude $A_{\rm s}$, the spectral tilt $n_{\rm s}$, the sound horizon $\theta_{*}$, the optical depth $\tau$, baryon density $\Omega_{\rm b}h^{2}$ and dark matter density $\Omega_{\rm dm}h^{2}$. For \LeDM{} we also vary the dark matter EoS today $w_{\rm dm,0}$. In \LeDM{} there is a strong degeneracy between $\Omega_{\rm dm}$ and $w_{\rm dm,0}$, which makes sampling the posterior extremely difficult. To remove this degeneracy and improve the efficiency of the MCMC sampling, we instead sample the initial dark matter density, i.e.~$\Omega_{\rm dm}^{\rm init}$ which is related to $\Omega_{\rm dm}$ by the relation $\Omega_{\rm dm} = \Omega_{\rm dm}^{\rm init}/W_{0}$, for \LCDM{} the two are equivalent since $W_{0}=1$. Note, the addition of the Type Ia absolute magnitude from SH0ES means we have to additionally sample $M_{B}$ as a sampled parameter. In Fig.~\ref{fig_params} we show the posterior of the sampled \LeDM{} parameters, showing that with the exception of $w_{\rm dm,0}$ the posteriors are generally quite Gaussian. A comparison to \LCDM{} shows the constraints on these parameters are close to identical for P18, since the high redshift CMB measurements are unable to constrain the late evolution of dark matter's EoS. To ensure the MCMC chains have converged we make use of \texttt{Cobaya}'s inbuilt measure of the $R-1$ Gelman-Rubin statistics \cite{Lewis2013}. The chains are assumed to be converged once $R-1<0.01$.

\begin{table*}
	\small
	\caption{Constraints on \LCDM{} and \LeDM{} parameters and statistics are shown for P18, P18+SN, P18+SN+$M_{B}$ and P18+SN+$M_{B}$+BAO.  In the top section we show the constraints on the base \LCDM{} parameters $A_{\rm s}$, $n_{\rm s}$, $\theta_{*}$, $\tau$, $\Omega_{\rm b}h^{2}$ and $\Omega_{\rm dm}^{\rm init}h^{2}$. For \LeDM{} we provide constraints on $w_{\rm dm,0}$ and $M_{B}$ when including constraints from SH0ES on the SN Type Ia absolute magnitude \cite{RiessMb2021}. In the second section we provide constraints on the derived parameters $H_{0}$, $\sigma_{8}$ and $S_{8}$. In the third section we provide estimations of the significance with respect to local measurements of $H_{0}$ \cite{Riess2021} and weak lensing measurements of $S_{8}$ from \citep{Kids2021Asgari} and  \citep{DESKids2023} in brackets. In the last section we provide comparisons of the $\chi^{2}$ and AIC for model selection, showing that although \LeDM{} provides a better fit to the data, there remains a marginal preference for \LCDM{} (based on the AIC). \label{tab_costraints2}}
	\begin{center}
		\begin{tabular}{c|cc|cc|cc|cc}
			\hline
			\T\B& \multicolumn{2}{c|}{P18}  &  \multicolumn{2}{c|}{...$\rm +SN$} &  \multicolumn{2}{c|}{...$+M_{B}$} &  \multicolumn{2}{c}{...$\rm +BAO$} \\ \cline{2-9}
			\T\B & \multicolumn{1}{c|}{\LCDM{}} & \multicolumn{1}{c|}{\LeDM{}} & \multicolumn{1}{c|}{\LCDM{}}  & \multicolumn{1}{c|}{\LeDM{}} & \multicolumn{1}{c|}{\LCDM{}} & \multicolumn{1}{c|}{\LeDM{}} & \multicolumn{1}{c|}{\LCDM{}}  & \multicolumn{1}{c}{\LeDM{}} \\ \hline
			\T\B $\ln(10^{10}A_{\rm s})$ & $3.045^{+0.029}_{-0.028}$ & $3.045^{+0.030}_{-0.028}$ & $3.045\pm0.028$ & $3.046^{+0.029}_{-0.028}$ & $3.048^{+0.030}_{-0.028}$ & $3.050^{+0.031}_{-0.028}$ & $3.048^{+0.029}_{-0.027}$ & $3.051^{+0.030}_{-0.028}$ \\
			\T\multirow{2}{*}{$n_{\rm s}$} &  $0.9659$ & $0.9660$ & $0.9658$ & $0.9665$ & $0.9672$ & $0.9688$ & $0.9675$ & $0.9692$ \\
			\B &  $^{+0.0080}_{-0.0079}$ & $\pm0.0082$ & $^{+0.0078}_{-0.0079}$ & $^{+0.0082}_{-0.0080}$ & $^{+0.0082}_{-0.0081}$ & $^{+0.0080}_{-0.0081}$ & $\pm0.0075$ & $^{+0.0071}_{-0.0074}$ \\
			\T \multirow{2}{*}{$100\,\theta_{\rm *}$} & $1.04196$ & $1.04196$ & $1.04196$ & $1.04198$ & $1.04203$ & $1.04210$ & $1.04204$ & $1.04212$ \\
			\B & $\pm0.00056$ & $^{+0.00057}_{-0.00060}$ & $\pm0.00055$ & $\pm0.00056$ & $^{+0.00058}_{-0.00057}$ & $^{+0.00057}_{-0.00058}$ & $^{+0.00057}_{-0.00056}$ & $^{+0.00055}_{-0.00056}$ \\
			\T\B $\tau$ &  $0.055^{+0.015}_{-0.014}$ & $0.055^{+0.016}_{-0.014}$ & $0.055^{+0.015}_{-0.014}$ & $0.055^{+0.015}_{-0.014}$ & $0.056^{+0.016}_{-0.014}$ & $0.058^{+0.016}_{-0.015}$ & $0.057^{+0.015}_{-0.014}$ & $0.059^{+0.015}_{-0.014}$ \\
			\multirow{2}{*}{$\Omega_{\rm b}h^{2}$} & $0.02239$ & $0.02240$ & $0.02239$ & $0.02241$ & $0.02245$ & $0.02250$ & $0.02245$ & $0.02251$ \\
			& $\pm 0.00029$ & $\pm 0.00029$ & $\pm 0.00028$ & $^{+0.00029}_{-0.00028}$ & $^{+0.00029}_{-0.00028}$ & $\pm 0.00028$ & $^{+0.00028}_{-0.00027}$ & $^{+0.00027}_{-0.00026}$ \\
			\T \multirow{2}{*}{$\Omega_{\rm dm}^{\rm init}h^{2}$} &  $0.1199$ & $0.1198$ & $0.1199$ & $0.1196$ & $0.1194$ & $0.1187$ & $0.1193$ & $0.1185$\\
			\B  &  $^{+0.0023}_{-0.0024}$ & $^{+0.0024}_{-0.0025}$ & $\pm0.0023$ & $^{+0.0022}_{-0.0023}$ & $^{+0.0023}_{-0.0024}$ & $^{+0.0023}_{-0.0022}$ & $\pm0.0020$ & $^{+0.0018}_{-0.0017}$\\		\T\B $w_{\rm dm,0}$ &  n/a & --- & n/a & $> -0.439$ & n/a  & $-0.31^{+0.30}_{-0.25}$ & n/a & $> -0.462$ \\
			\T\B $M_{B}$ & n/a & n/a & n/a & n/a & $-19.411\pm0.028$ & $-19.386^{+0.039}_{-0.038}$ &
			$-19.409^{+0.023}_{-0.022}$ & $-19.393^{+0.030}_{-0.028}$ \\ \hline
			\T\B Derived & & & & & & & & \\ \hline
			\T\B $H_{0}$ &  $67.4\pm1.1$ & $69.7^{+4.3}_{-3.0}$ & $67.5^{+1.0}_{-0.99}$ & $68.3^{+1.6}_{-1.5}$ & $67.9\pm1.0$ & $69.2^{+1.8}_{-1.7}$ & $67.99^{+0.81}_{-0.80}$ & $68.8^{+1.3}_{-1.2}$ \\
			\T\B $\sigma_{8}$ &  $0.811\pm0.012$ & $0.59^{+0.21}_{-0.22}$ & $0.810\pm0.012$ & $0.69^{+0.12}_{-0.13}$ & $0.809\pm0.012$ & $0.64^{+0.15}_{-0.14}$ & $0.809\pm0.012$ & $0.67\pm0.13$ \\
			\T\B $S_{8}$ &  $0.829^{+0.025}_{-0.026}$  & $0.59^{+0.23}_{-0.25}$  & $0.827\pm0.024$  & $0.70^{+0.13}_{-0.14}$  & $0.818\pm0.024$  & $0.63^{+0.16}_{-0.15}$  & $0.816\pm0.020$  & $0.67^{+0.14}_{-0.13}$  \\ \hline
			\T\B Tensions & & & & & & & & \\ \hline
			\T $H_{0}$ & $5.04\sigma$ & $1.73\sigma$ & $5.02\sigma$ & $3.86\sigma$ & n/a & n/a & n/a & n/a \\
			\T  \multirow{2}{*}{$S_{8}$} & $2.94\sigma$ & $1.13\sigma$ & $2.93\sigma$ & $0.83\sigma$ & $2.5\sigma$& $1.53\sigma$ & $2.54\sigma$ & $1.21\sigma$ \\
			\B & $(1.89\sigma)$ & $(1.28\sigma)$ & $(1.85\sigma)$ & $(1.12\sigma)$ & $(1.4\sigma)$ & $(1.8\sigma)$ & $(1.38\sigma)$ & $(1.52\sigma)$\\ \hline
			\T Model & & & & & & & & \\
			\B Selection & & & & & & & & \\\hline
			\T\B $\chi^{2}$ & $1012.02$ & $1012.24$ & $2047.35$ & $2046.87$ & $2057.42$ & $2054.85$ & $2063.32$ & $2062$ \\
			\T\B $\Delta \chi^{2}$ & $0$ & $0.22$ & $0$ & $-0.48$ & $0$ & $-2.57$ & $0$ & $-1.32$ \\
			\T\B AIC & $1013.15$ & $1015.34$ & $2048.35$ & $2049.95$ & $2061.88$ & $2062.18 $ & $2067.69$ & $2068.42$ \\
			\T\B $\Delta$AIC & $0$ & $2.19$ & $0$ & $1.6$ & $0$ & $0.31$ & $0$ &  $0.72$ \\ \hline
		\end{tabular}
	\end{center}
\end{table*}

Defining $S_{8}$ for \LeDM{} is rather challenging since the general definition $S_{8}=\sigma_{8}\sqrt{0.3/\Omega_{\rm m}}$ in the literature generally assumes the contribution of cold dark matter to $\Omega_{\rm m}$. Assuming that this functional form is really trying to capture an amplitude to the power spectra and define this in relation to the fiducial \LCDM{} constraints where $\Omega_{\rm m}\simeq0.3$ the most sensible definition of $S_{8}$ will use $\Omega_{\rm dm}^{\rm init}$ since this sets the shape and amplitude of the initial power spectra and using $\sigma_{8}$ which captures \LeDM{}'s dampening effect. Therefore, we define
\begin{equation}
	S_{8} = \sigma_{8}\sqrt{\Omega_{\rm m}^{\rm init}/0.3}
\end{equation}
where $\Omega_{\rm m}^{\rm init} = \Omega_{\rm dm}^{\rm init} + \Omega_{\rm b} + \Omega_{\nu}^{\rm NR} $ and $\Omega_{\nu}^{\rm NR}$ is the contribution from non-relativistic neutrinos.

In Table~\ref{tab_costraints2} we show the constraints on the base and derived parameters for \LCDM{} and \LeDM{}. For model comparison, we provide the $\chi^{2}$, Akaike Information Criterion (AIC; \citep{Akaike1974}) and the significance of the Hubble, $M_{B}$ and $S_{8}$ tensions; comparing to $H_{0}=73.3\pm1.04$ \cite{Riess2021}, $M_{B}=-19.244\pm0.020$ \cite{RiessMb2021} and $S_{8}=0.766\pm0.017$ \cite{Kids2021} ($S_{8}=0.790\pm0.016$ in brackets; \cite{DESKids2023}) respectively.

\section{Results}

\subsection{Constraints on \LeDM{}}

\begin{figure}
	\includegraphics[width=\columnwidth]{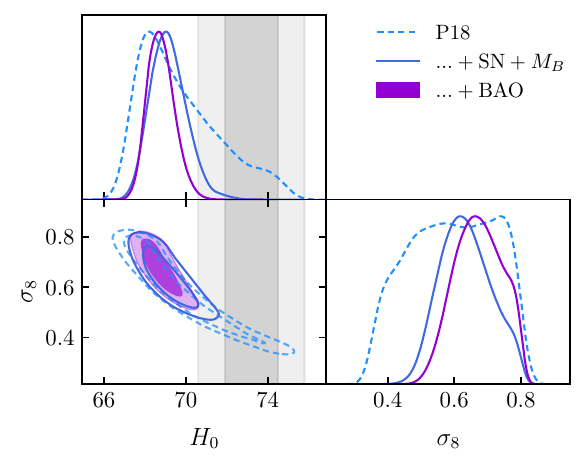}
	\caption{Constraints on the derived parameters of $H_{0}$ and \sig{} are shown for \LeDM{}. These are constrained with Planck CMB measurements (dashed blue contours), the addition of Pantheon type Ia supernovae and Cepheid constraints on $M_{B}$ (SN+$M_{B}$; solid blue contours) and BAO (purple shaded contours). The contours indicate regions of 68\% and 95\% confidence intervals. Constraints on \LeDM{} provide higher values of $H_{0}$ and lower values of \sig{} easing the $H_{0}$ and $S_{8}$ tensions. \label{fig_H0_sigma8}}
\end{figure}

We constrain \LeDM{} parameters (standard \LCDM{} parameters plus the current dark matter EoS $w_{\rm dm, 0}$) using measurements from Planck CMB (referred to as P18, \cite{Aghanim2018, Aghanim2019}), Type Ia supernovae from Pantheon \cite{Scolnic2017} (referred to as SN), SH0ES constraints on the absolute magnitude of Type Ia supernovae $M_{B}$ from Cepheids \cite{RiessMb2021, Efstathiou2021, Camarena2021} (referred to as SN+$M_{B}$), baryonic acoustic oscillations and redshift space distortions from 6dF, SDSS DR7 and BOSS \citep{Beutler2012, Ross2014, Alam2016} (referred to as BAO). Constraints on cosmological parameters for \LCDM{} and \LeDM{} are provided in Table~\ref{tab_costraints2}. In Fig.~\ref{fig_H0_sigma8} we highlight the constraints on the derived quantities of $H_{0}$ and \sig{}, showing that \LeDM{} allows for slightly larger values of $H_{0}$ ($\sim 68.3$ for \LeDM{} compared to $\sim 67.5$ for \LCDM{} from P18+SN constraints), easing tensions with measurements of SH0ES $M_{B}$ \cite{RiessMb2021} from $\sim5\sigma$ to $\sim3\sigma$. The larger $H_{0}$ constraints allow for lower values of \sig{} ($\sim0.69$ for \LeDM{} compared to $\sim 0.81$ for \LCDM{} from constraints P18+SN constraints), a result which eases tensions with weak lensing measurements \cite{Kids2021Asgari, DES2022, DESKids2023, HSC2023} from $\sim3\sigma$ to $\sim1\sigma$.

In Table~\ref{tab_costraints2} we provide the $\chi^{2}$ and AIC for model selection. Comparisons of the $\chi^{2}$ show that combinations of P18 with SN, SN+$M_{B}$ and SN+$M_{B}$+BAO are better fitted with a \LeDM{} model however the AIC penalises the necessity for the $w_{\rm dm,0}$ parameter, showing a marginal preference for \LCDM{}. In this Table we show the significance of tensions between the constraints on the Hubble constant (comparing to $H_{0}=73.3\pm1.04$ \citep{Riess2021}) and $S_{8}$ (comparing to $S_{8}=0.766\pm0.017$ \citep{Kids2021} and in brackets to $S_{8}=0.790\pm0.016$ \citep{DESKids2023}). Here the significance is obtained by computing the difference between the best-fit values and dividing by the joint errors added in quadrature. The significance of the $H_{0}$ tension are reduced from 4-5$\sigma$ to $\sim3.5\sigma$ while the $S_{8}$ tension is reduced from $\sim3\sigma$ to $\sim1\sigma$. The constraints on the dark matter EoS remain consistent with \LCDM{} $w_{\rm dm,0} > -0.462$ (for P18+SN+$M_{B}$+BAO), although the maximum likelihood peaks at $w_{\rm dm,0} \sim -0.2$.

\begin{figure*}
	\includegraphics[width=\textwidth]{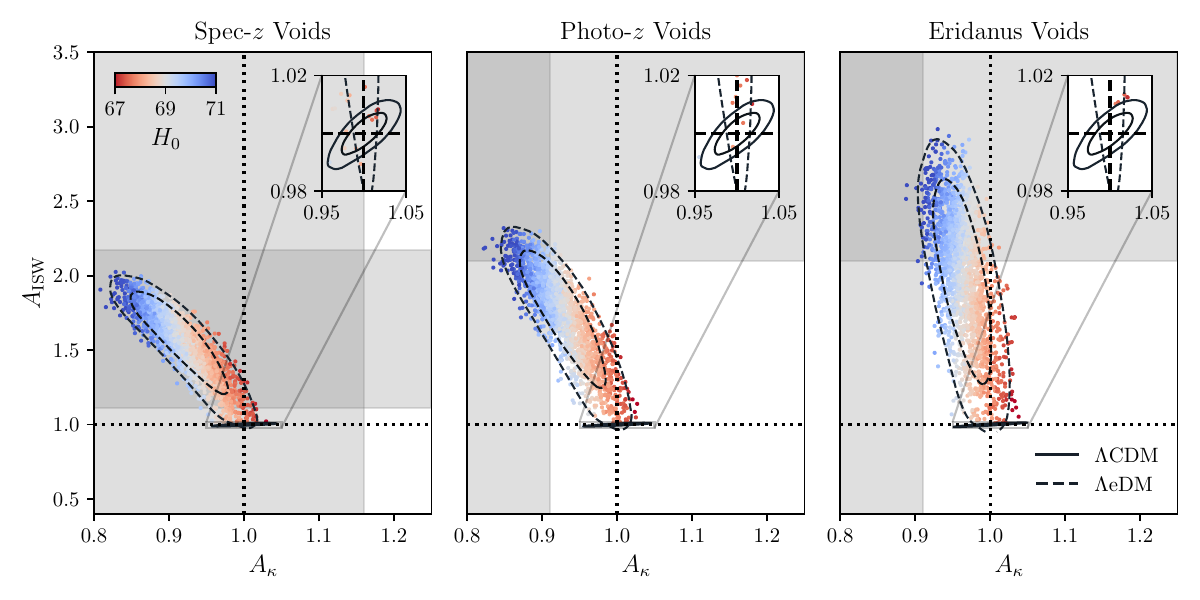}
	\caption{The amplitude of the ISW ($A_{\rm ISW}$) and lensing convergence ($A_{\kappa}$) for three void types are sampled from parameter constraints on \LCDM{} (solid black) and \LeDM{} (dashed black) from P18+SN+$M_{B}$+BAO. These are compared to the amplitude obtained from the best-fit constraints on \LCDM{}. The Hubble constant value for the sampled points in \LeDM{} are indicated with a red-blue color bar. The grey horizontal and vertical bands show measurements from observations (see Table~\ref{tab_voids}). Constraints from \LCDM{} show the amplitudes should be very close to one. For \LeDM{} a larger $H_{0}$ strongly correlates to a much larger $A_{\rm ISW}$ and smaller $A_{\kappa}$. \label{fig_AiswAkappa}}
\end{figure*}

\subsection{ISW and Lensing Predictions}

In Fig.~\ref{fig_AiswAkappa} we sample the chains from the \texttt{Cobaya} MCMC for \LCDM{} and \LeDM{} from P18+SN+$M_{B}$+BAO and display the amplitude of the ISW ($A_{\rm ISW}$) and lensing ($A_{\kappa}$) for the three void types with respect to the best-fit \LCDM{} model. We show here that \LCDM{} makes very tight predictions for what this should be, with little dependence on the void types. On the other hand \LeDM{} allows for much larger $A_{\rm ISW}$ and smaller $A_{\kappa}$. The Hubble constant for the sampled points is shown with a red-blue color bar, indicating that in \LeDM{} a higher $H_{0}$ is met with larger $A_{\rm ISW}$ and smaller $A_{\kappa}$, in better agreement with observations. A redshift evolution can be seen in the profiles, voids at lower redshift exhibit a larger $A_{\rm ISW}$ but weaker $A_{\kappa}$ with respect to \LCDM{} for high-$H_{0}$.

\begin{table}
	\caption{Predictions for the ISW $A_{\rm ISW}$ and lensing convergence $A_{\kappa}$ for the Spec-$z$, Photo-$z$ and Eridanus voids from constraints of P18+SN+$M_{B}$+BAO on \LCDM{} and \LeDM{}. For \LCDM{} the predictions are very close to $~1$, while \LeDM{} allows for $A_{\rm ISW}$ to be a factor of up to $2$ times larger than \LCDM{} and $A_{\kappa}$ a factor of up to $0.9$ times smaller than \LCDM{}. \label{tab_pred_voids}}
	\begin{center}
		\begin{tabular}{l|cc|cc}
			\hline
			\T\B \multirow{2}{*}{Void} & \multicolumn{2}{c|}{$A_{\rm ISW}$} & \multicolumn{2}{c}{$A_{\kappa}$} \\ \cline{2-5}
			\T\B & \multicolumn{1}{c|}{\LCDM{}} & \multicolumn{1}{c|}{\LeDM{}} & \multicolumn{1}{c|}{\LCDM{}} & \multicolumn{1}{c}{\LeDM{}} \\
			\hline
			\T\B Spec-$z$ & $1^{+0.0087}_{-0.0093}$ & $1.55^{+0.31}_{-0.49}$ & $1_{-0.032}^{+0.033}$ &
			$0.92_{-0.072}^{+0.079}$ \\
			\T\B Photo-$z$ & $1^{+0.0096}_{-0.01}$ & $1.71^{+0.44}_{-0.64}$ & $1^{+0.035}_{-0.034}$ & $0.93^{+0.069}_{-0.065}$ \\
			\T\B  Eridanus & $1\pm0.012$ & $1.97^{+0.71}_{-0.86}$ & $1\pm0.037$ & $0.96^{+0.048}_{-0.046}$ \\
			\hline
		\end{tabular}
	\end{center}
\end{table}

In Table~\ref{tab_pred_voids} we make predictions for the void ISW and lensing amplitude for \LeDM{} with respect to \LCDM{} based on constraints using P18+SN+$M_{B}$+BAO. These values are in excellent agreement with current observations and suggest the $S_{8}$ tensions and void-ISW anomaly can be solved and the $H_{0}$ tension alleviated with new physics in dark matter (or a unified interacting dark sector) that is being captured by the phenomenological \LeDM{} model.

\section{Discussions}

In this paper we introduce \LeDM{}, a phenomenological model of dark matter with a non-zero EoS at late times. We show that \LeDM{} is able to simultaneously ease the $H_{0}$ and $S_{8}$ tensions while also providing an explanation for the void-ISW anomaly. This is the first model, to our knowledge, that can consistently explain all three with a single modification.

Previous solutions to the $H_{0}$ tension (such as early dark energy \cite{EDE}) provide a high $H_{0}$ at the cost of greater clustering and therefore larger \sig{} -- putting them in conflict with the $S_{8}$ tension \citep{Vagnozzi2021,Escudero2022}. \LeDM{}, on the other hand, does not require changing the physical densities of dark matter at last scattering but instead imposes a late change to the evolution of dark matter. Since dark matter has never been directly observed, it is reasonable to suspect the simple assumptions of cold dark matter may be incomplete. Best fit constraints on \LeDM{} imply dark matter's EoS is negative at late times, strongly implying interactions between dark matter and dark energy \citep{Cesare2022}, which have been shown to alleviate both the $H_{0}$ and $S_{8}$ tensions \citep{Pourtsidou2016, Kumar2016,  Kumar2019, DiValentino2020, Lucca2020, Poulin2022}; if this is proven to be true this will have a profound impact on fundamental physics.

However \LeDM{} does not only alleviate the $H_{0}$ and $S_{8}$ tensions, it also provides an explanation to a long-standing anomaly -- the void-ISW anomaly, while remaining consistent with void lensing measurements. This allows the model to be tested in the near future, including testing other shapes and parameterisations for the EoS, from galaxy surveys such as Euclid\footnote{\href{https://www.euclid-ec.org/}{https://www.euclid-ec.org/}} and LSST\footnote{\href{https://www.lsst.org/}{https://www.lsst.org/}}. So far, observations appear to be consistent with constraints to \LeDM{}, with the ISW of voids predicted to be roughly a factor of $\sim2$ and lensing to be a factor of $\sim0.9$ with respect to \LCDM{}. Future measurements of void ISW and lensing, and other probes (such as galaxy cluster abundances, galaxy lensing and CMB temperature auto and cross-correlations) will allow us to provide tighter constraints on \LeDM{} and help determine whether new physics in dark matter (or an interacting dark sector) is the cause of the $H_{0}$ and $S_{8}$ tension.

\section{Acknowledgements}%
We thank Enrique Gazta\~{n}aga, Ofer Lahav, Constance Mahony, Adi Nusser and Joe Silk for providing useful suggestions and insightful discussions in preparation of this paper. Cosmological constraints on \LeDM{} were obtained using the MCMC \cite{Lewis2002, Neal2005, Lewis2013} package \texttt{Cobaya} \cite{CobayaSoftware2019, CobayaPaper2021}. This work is supported by the National Science Center, Poland, under grant agreements no: 2018/30/E/ST9/00698 and 2018/31/G/ST9/03388. MJ is supported by the Polish Ministry of Science and Higher Education through grant DIR/WK/2018/12. MB is supported by the Polish National Science Center through grants no. 2020/38/E/ST9/00395 and 2020/39/B/ST9/03494, and by the Polish Ministry of Science and Higher Education through grant DIR/WK/2018/12.

\bibliography{biblio}

\providecommand{\noopsort}[1]{}\providecommand{\singleletter}[1]{#1}%
\begin{thebibliography}{71}%
\makeatletter
\providecommand \@ifxundefined [1]{%
 \@ifx{#1\undefined}
}%
\providecommand \@ifnum [1]{%
 \ifnum #1\expandafter \@firstoftwo
 \else \expandafter \@secondoftwo
 \fi
}%
\providecommand \@ifx [1]{%
 \ifx #1\expandafter \@firstoftwo
 \else \expandafter \@secondoftwo
 \fi
}%
\providecommand \natexlab [1]{#1}%
\providecommand \enquote  [1]{``#1''}%
\providecommand \bibnamefont  [1]{#1}%
\providecommand \bibfnamefont [1]{#1}%
\providecommand \citenamefont [1]{#1}%
\providecommand \href@noop [0]{\@secondoftwo}%
\providecommand \href [0]{\begingroup \@sanitize@url \@href}%
\providecommand \@href[1]{\@@startlink{#1}\@@href}%
\providecommand \@@href[1]{\endgroup#1\@@endlink}%
\providecommand \@sanitize@url [0]{\catcode `\\12\catcode `\$12\catcode
  `\&12\catcode `\#12\catcode `\^12\catcode `\_12\catcode `\%12\relax}%
\providecommand \@@startlink[1]{}%
\providecommand \@@endlink[0]{}%
\providecommand \url  [0]{\begingroup\@sanitize@url \@url }%
\providecommand \@url [1]{\endgroup\@href {#1}{\urlprefix }}%
\providecommand \urlprefix  [0]{URL }%
\providecommand \Eprint [0]{\href }%
\providecommand \doibase [0]{https://doi.org/}%
\providecommand \selectlanguage [0]{\@gobble}%
\providecommand \bibinfo  [0]{\@secondoftwo}%
\providecommand \bibfield  [0]{\@secondoftwo}%
\providecommand \translation [1]{[#1]}%
\providecommand \BibitemOpen [0]{}%
\providecommand \bibitemStop [0]{}%
\providecommand \bibitemNoStop [0]{.\EOS\space}%
\providecommand \EOS [0]{\spacefactor3000\relax}%
\providecommand \BibitemShut  [1]{\csname bibitem#1\endcsname}%
\let\auto@bib@innerbib\@empty
\bibitem [{\citenamefont {{Planck Collaboration}}\ \emph
  {et~al.}(2020{\natexlab{a}})\citenamefont {{Planck Collaboration}},
  \citenamefont {{Aghanim}}, \citenamefont {{Akrami}}, \citenamefont
  {{Ashdown}}, \citenamefont {{Aumont}}, \citenamefont {{Baccigalupi}},
  \citenamefont {{Ballardini}}, \citenamefont {{Banday}}, \citenamefont
  {{Barreiro}}, \citenamefont {{Bartolo}}, \citenamefont {{Basak}},
  \citenamefont {{Battye}}, \citenamefont {{Benabed}}, \citenamefont
  {{Bernard}}, \citenamefont {{Bersanelli}} \emph {et~al.}}]{Planck2018}%
  \BibitemOpen
  \bibfield  {author} {\bibinfo {author} {\bibnamefont {{Planck
  Collaboration}}}, \bibinfo {author} {\bibfnamefont {N.}~\bibnamefont
  {{Aghanim}}}, \bibinfo {author} {\bibfnamefont {Y.}~\bibnamefont {{Akrami}}},
  \bibinfo {author} {\bibfnamefont {M.}~\bibnamefont {{Ashdown}}}, \bibinfo
  {author} {\bibfnamefont {J.}~\bibnamefont {{Aumont}}}, \bibinfo {author}
  {\bibfnamefont {C.}~\bibnamefont {{Baccigalupi}}}, \bibinfo {author}
  {\bibfnamefont {M.}~\bibnamefont {{Ballardini}}}, \bibinfo {author}
  {\bibfnamefont {A.~J.}\ \bibnamefont {{Banday}}}, \bibinfo {author}
  {\bibfnamefont {R.~B.}\ \bibnamefont {{Barreiro}}}, \bibinfo {author}
  {\bibfnamefont {N.}~\bibnamefont {{Bartolo}}}, \bibinfo {author}
  {\bibfnamefont {S.}~\bibnamefont {{Basak}}}, \bibinfo {author} {\bibfnamefont
  {R.}~\bibnamefont {{Battye}}}, \bibinfo {author} {\bibfnamefont
  {K.}~\bibnamefont {{Benabed}}}, \bibinfo {author} {\bibfnamefont {J.~P.}\
  \bibnamefont {{Bernard}}}, \bibinfo {author} {\bibfnamefont {M.}~\bibnamefont
  {{Bersanelli}}}, \emph {et~al.},\ }\bibfield  {title} {\bibinfo {title}
  {{Planck 2018 results. VI. Cosmological parameters}},\ }\href
  {https://doi.org/10.1051/0004-6361/201833910} {\bibfield  {journal} {\bibinfo
   {journal} {A\&A}\ }\textbf {\bibinfo {volume} {641}},\ \bibinfo {eid} {A6}
  (\bibinfo {year} {2020}{\natexlab{a}})},\ \Eprint
  {https://arxiv.org/abs/1807.06209} {arXiv:1807.06209 [astro-ph.CO]}
  \BibitemShut {NoStop}%
\bibitem [{\citenamefont {{Riess}}\ \emph {et~al.}(2016)\citenamefont
  {{Riess}}, \citenamefont {{Macri}}, \citenamefont {{Hoffmann}}, \citenamefont
  {{Scolnic}}, \citenamefont {{Casertano}}, \citenamefont {{Filippenko}},
  \citenamefont {{Tucker}}, \citenamefont {{Reid}}, \citenamefont {{Jones}},
  \citenamefont {{Silverman}}, \citenamefont {{Chornock}}, \citenamefont
  {{Challis}}, \citenamefont {{Yuan}}, \citenamefont {{Brown}},\ and\
  \citenamefont {{Foley}}}]{Riess2016}%
  \BibitemOpen
  \bibfield  {author} {\bibinfo {author} {\bibfnamefont {A.~G.}\ \bibnamefont
  {{Riess}}}, \bibinfo {author} {\bibfnamefont {L.~M.}\ \bibnamefont
  {{Macri}}}, \bibinfo {author} {\bibfnamefont {S.~L.}\ \bibnamefont
  {{Hoffmann}}}, \bibinfo {author} {\bibfnamefont {D.}~\bibnamefont
  {{Scolnic}}}, \bibinfo {author} {\bibfnamefont {S.}~\bibnamefont
  {{Casertano}}}, \bibinfo {author} {\bibfnamefont {A.~V.}\ \bibnamefont
  {{Filippenko}}}, \bibinfo {author} {\bibfnamefont {B.~E.}\ \bibnamefont
  {{Tucker}}}, \bibinfo {author} {\bibfnamefont {M.~J.}\ \bibnamefont
  {{Reid}}}, \bibinfo {author} {\bibfnamefont {D.~O.}\ \bibnamefont {{Jones}}},
  \bibinfo {author} {\bibfnamefont {J.~M.}\ \bibnamefont {{Silverman}}},
  \bibinfo {author} {\bibfnamefont {R.}~\bibnamefont {{Chornock}}}, \bibinfo
  {author} {\bibfnamefont {P.}~\bibnamefont {{Challis}}}, \bibinfo {author}
  {\bibfnamefont {W.}~\bibnamefont {{Yuan}}}, \bibinfo {author} {\bibfnamefont
  {P.~J.}\ \bibnamefont {{Brown}}},\ and\ \bibinfo {author} {\bibfnamefont
  {R.~J.}\ \bibnamefont {{Foley}}},\ }\bibfield  {title} {\bibinfo {title} {{A
  2.4\% Determination of the Local Value of the Hubble Constant}},\ }\href
  {https://doi.org/10.3847/0004-637X/826/1/56} {\bibfield  {journal} {\bibinfo
  {journal} {\apj}\ }\textbf {\bibinfo {volume} {826}},\ \bibinfo {eid} {56}
  (\bibinfo {year} {2016})},\ \Eprint {https://arxiv.org/abs/1604.01424}
  {arXiv:1604.01424 [astro-ph.CO]} \BibitemShut {NoStop}%
\bibitem [{\citenamefont {{Asgari}}\ \emph {et~al.}(2021)\citenamefont
  {{Asgari}}, \citenamefont {{Lin}}, \citenamefont {{Joachimi}}, \citenamefont
  {{Giblin}}, \citenamefont {{Heymans}}, \citenamefont {{Hildebrandt}},
  \citenamefont {{Kannawadi}}, \citenamefont {{St{\"o}lzner}}, \citenamefont
  {{Tr{\"o}ster}}, \citenamefont {{van den Busch}}, \citenamefont {{Wright}},
  \citenamefont {{Bilicki}}, \citenamefont {{Blake}}, \citenamefont {{de
  Jong}}, \citenamefont {{Dvornik}}, \citenamefont {{Erben}}, \citenamefont
  {{Getman}}, \citenamefont {{Hoekstra}}, \citenamefont {{K{\"o}hlinger}},
  \citenamefont {{Kuijken}}, \citenamefont {{Miller}}, \citenamefont
  {{Radovich}}, \citenamefont {{Schneider}}, \citenamefont {{Shan}},\ and\
  \citenamefont {{Valentijn}}}]{Kids2021Asgari}%
  \BibitemOpen
  \bibfield  {author} {\bibinfo {author} {\bibfnamefont {M.}~\bibnamefont
  {{Asgari}}}, \bibinfo {author} {\bibfnamefont {C.-A.}\ \bibnamefont {{Lin}}},
  \bibinfo {author} {\bibfnamefont {B.}~\bibnamefont {{Joachimi}}}, \bibinfo
  {author} {\bibfnamefont {B.}~\bibnamefont {{Giblin}}}, \bibinfo {author}
  {\bibfnamefont {C.}~\bibnamefont {{Heymans}}}, \bibinfo {author}
  {\bibfnamefont {H.}~\bibnamefont {{Hildebrandt}}}, \bibinfo {author}
  {\bibfnamefont {A.}~\bibnamefont {{Kannawadi}}}, \bibinfo {author}
  {\bibfnamefont {B.}~\bibnamefont {{St{\"o}lzner}}}, \bibinfo {author}
  {\bibfnamefont {T.}~\bibnamefont {{Tr{\"o}ster}}}, \bibinfo {author}
  {\bibfnamefont {J.~L.}\ \bibnamefont {{van den Busch}}}, \bibinfo {author}
  {\bibfnamefont {A.~H.}\ \bibnamefont {{Wright}}}, \bibinfo {author}
  {\bibfnamefont {M.}~\bibnamefont {{Bilicki}}}, \bibinfo {author}
  {\bibfnamefont {C.}~\bibnamefont {{Blake}}}, \bibinfo {author} {\bibfnamefont
  {J.}~\bibnamefont {{de Jong}}}, \bibinfo {author} {\bibfnamefont
  {A.}~\bibnamefont {{Dvornik}}}, \bibinfo {author} {\bibfnamefont
  {T.}~\bibnamefont {{Erben}}}, \bibinfo {author} {\bibfnamefont
  {F.}~\bibnamefont {{Getman}}}, \bibinfo {author} {\bibfnamefont
  {H.}~\bibnamefont {{Hoekstra}}}, \bibinfo {author} {\bibfnamefont
  {F.}~\bibnamefont {{K{\"o}hlinger}}}, \bibinfo {author} {\bibfnamefont
  {K.}~\bibnamefont {{Kuijken}}}, \bibinfo {author} {\bibfnamefont
  {L.}~\bibnamefont {{Miller}}}, \bibinfo {author} {\bibfnamefont
  {M.}~\bibnamefont {{Radovich}}}, \bibinfo {author} {\bibfnamefont
  {P.}~\bibnamefont {{Schneider}}}, \bibinfo {author} {\bibfnamefont
  {H.}~\bibnamefont {{Shan}}},\ and\ \bibinfo {author} {\bibfnamefont
  {E.}~\bibnamefont {{Valentijn}}},\ }\bibfield  {title} {\bibinfo {title}
  {{KiDS-1000 cosmology: Cosmic shear constraints and comparison between two
  point statistics}},\ }\href {https://doi.org/10.1051/0004-6361/202039070}
  {\bibfield  {journal} {\bibinfo  {journal} {\aap}\ }\textbf {\bibinfo
  {volume} {645}},\ \bibinfo {eid} {A104} (\bibinfo {year} {2021})},\ \Eprint
  {https://arxiv.org/abs/2007.15633} {arXiv:2007.15633 [astro-ph.CO]}
  \BibitemShut {NoStop}%
\bibitem [{\citenamefont {{Abbott}}\ \emph {et~al.}(2022)\citenamefont
  {{Abbott}}, \citenamefont {{Aguena}}, \citenamefont {{Alarcon}},
  \citenamefont {{Allam}}, \citenamefont {{Alves}}, \citenamefont {{Amon}},
  \citenamefont {{Andrade-Oliveira}}, \citenamefont {{Annis}}, \citenamefont
  {{Avila}}, \citenamefont {{Bacon}}, \citenamefont {{Baxter}}, \citenamefont
  {{Bechtol}}, \citenamefont {{Becker}}, \citenamefont {{Bernstein}},
  \citenamefont {{Bhargava}} \emph {et~al.}}]{DES2022}%
  \BibitemOpen
  \bibfield  {author} {\bibinfo {author} {\bibfnamefont {T.~M.~C.}\
  \bibnamefont {{Abbott}}}, \bibinfo {author} {\bibfnamefont {M.}~\bibnamefont
  {{Aguena}}}, \bibinfo {author} {\bibfnamefont {A.}~\bibnamefont {{Alarcon}}},
  \bibinfo {author} {\bibfnamefont {S.}~\bibnamefont {{Allam}}}, \bibinfo
  {author} {\bibfnamefont {O.}~\bibnamefont {{Alves}}}, \bibinfo {author}
  {\bibfnamefont {A.}~\bibnamefont {{Amon}}}, \bibinfo {author} {\bibfnamefont
  {F.}~\bibnamefont {{Andrade-Oliveira}}}, \bibinfo {author} {\bibfnamefont
  {J.}~\bibnamefont {{Annis}}}, \bibinfo {author} {\bibfnamefont
  {S.}~\bibnamefont {{Avila}}}, \bibinfo {author} {\bibfnamefont
  {D.}~\bibnamefont {{Bacon}}}, \bibinfo {author} {\bibfnamefont
  {E.}~\bibnamefont {{Baxter}}}, \bibinfo {author} {\bibfnamefont
  {K.}~\bibnamefont {{Bechtol}}}, \bibinfo {author} {\bibfnamefont {M.~R.}\
  \bibnamefont {{Becker}}}, \bibinfo {author} {\bibfnamefont {G.~M.}\
  \bibnamefont {{Bernstein}}}, \bibinfo {author} {\bibfnamefont
  {S.}~\bibnamefont {{Bhargava}}}, \emph {et~al.},\ }\bibfield  {title}
  {\bibinfo {title} {{Dark Energy Survey Year 3 results: Cosmological
  constraints from galaxy clustering and weak lensing}},\ }\href
  {https://doi.org/10.1103/PhysRevD.105.023520} {\bibfield  {journal} {\bibinfo
   {journal} {\prd}\ }\textbf {\bibinfo {volume} {105}},\ \bibinfo {eid}
  {023520} (\bibinfo {year} {2022})},\ \Eprint
  {https://arxiv.org/abs/2105.13549} {arXiv:2105.13549 [astro-ph.CO]}
  \BibitemShut {NoStop}%
\bibitem [{\citenamefont {{Dark Energy Survey and Kilo-Degree Survey
  Collaboration}}\ \emph {et~al.}(2023)\citenamefont {{Dark Energy Survey and
  Kilo-Degree Survey Collaboration}}, \citenamefont {{Abbott}}, \citenamefont
  {{Aguena}}, \citenamefont {{Alarcon}}, \citenamefont {{Alves}}, \citenamefont
  {{Amon}}, \citenamefont {{Andrade-Oliveira}}, \citenamefont {{Asgari}},
  \citenamefont {{Avila}}, \citenamefont {{Bacon}}, \citenamefont {{Bechtol}},
  \citenamefont {{Becker}}, \citenamefont {{Bernstein}}, \citenamefont
  {{Bertin}}, \citenamefont {{Bilicki}} \emph {et~al.}}]{DESKids2023}%
  \BibitemOpen
  \bibfield  {author} {\bibinfo {author} {\bibnamefont {{Dark Energy Survey and
  Kilo-Degree Survey Collaboration}}}, \bibinfo {author} {\bibfnamefont
  {T.~M.~C.}\ \bibnamefont {{Abbott}}}, \bibinfo {author} {\bibfnamefont
  {M.}~\bibnamefont {{Aguena}}}, \bibinfo {author} {\bibfnamefont
  {A.}~\bibnamefont {{Alarcon}}}, \bibinfo {author} {\bibfnamefont
  {O.}~\bibnamefont {{Alves}}}, \bibinfo {author} {\bibfnamefont
  {A.}~\bibnamefont {{Amon}}}, \bibinfo {author} {\bibfnamefont
  {F.}~\bibnamefont {{Andrade-Oliveira}}}, \bibinfo {author} {\bibfnamefont
  {M.}~\bibnamefont {{Asgari}}}, \bibinfo {author} {\bibfnamefont
  {S.}~\bibnamefont {{Avila}}}, \bibinfo {author} {\bibfnamefont
  {D.}~\bibnamefont {{Bacon}}}, \bibinfo {author} {\bibfnamefont
  {K.}~\bibnamefont {{Bechtol}}}, \bibinfo {author} {\bibfnamefont {M.~R.}\
  \bibnamefont {{Becker}}}, \bibinfo {author} {\bibfnamefont {G.~M.}\
  \bibnamefont {{Bernstein}}}, \bibinfo {author} {\bibfnamefont
  {E.}~\bibnamefont {{Bertin}}}, \bibinfo {author} {\bibfnamefont
  {M.}~\bibnamefont {{Bilicki}}}, \emph {et~al.},\ }\bibfield  {title}
  {\bibinfo {title} {{DES Y3 + KiDS-1000: Consistent cosmology combining cosmic
  shear surveys}},\ }\href {https://doi.org/10.21105/astro.2305.17173}
  {\bibfield  {journal} {\bibinfo  {journal} {The Open Journal of
  Astrophysics}\ }\textbf {\bibinfo {volume} {6}},\ \bibinfo {eid} {36}
  (\bibinfo {year} {2023})},\ \Eprint {https://arxiv.org/abs/2305.17173}
  {arXiv:2305.17173 [astro-ph.CO]} \BibitemShut {NoStop}%
\bibitem [{\citenamefont {{Dalal}}\ \emph {et~al.}(2023)\citenamefont
  {{Dalal}}, \citenamefont {{Li}}, \citenamefont {{Nicola}}, \citenamefont
  {{Zuntz}}, \citenamefont {{Strauss}}, \citenamefont {{Sugiyama}},
  \citenamefont {{Zhang}}, \citenamefont {{Rau}}, \citenamefont {{Mandelbaum}},
  \citenamefont {{Takada}}, \citenamefont {{More}}, \citenamefont {{Miyatake}},
  \citenamefont {{Kannawadi}}, \citenamefont {{Shirasaki}}, \citenamefont
  {{Taniguchi}} \emph {et~al.}}]{HSC2023}%
  \BibitemOpen
  \bibfield  {author} {\bibinfo {author} {\bibfnamefont {R.}~\bibnamefont
  {{Dalal}}}, \bibinfo {author} {\bibfnamefont {X.}~\bibnamefont {{Li}}},
  \bibinfo {author} {\bibfnamefont {A.}~\bibnamefont {{Nicola}}}, \bibinfo
  {author} {\bibfnamefont {J.}~\bibnamefont {{Zuntz}}}, \bibinfo {author}
  {\bibfnamefont {M.~A.}\ \bibnamefont {{Strauss}}}, \bibinfo {author}
  {\bibfnamefont {S.}~\bibnamefont {{Sugiyama}}}, \bibinfo {author}
  {\bibfnamefont {T.}~\bibnamefont {{Zhang}}}, \bibinfo {author} {\bibfnamefont
  {M.~M.}\ \bibnamefont {{Rau}}}, \bibinfo {author} {\bibfnamefont
  {R.}~\bibnamefont {{Mandelbaum}}}, \bibinfo {author} {\bibfnamefont
  {M.}~\bibnamefont {{Takada}}}, \bibinfo {author} {\bibfnamefont
  {S.}~\bibnamefont {{More}}}, \bibinfo {author} {\bibfnamefont
  {H.}~\bibnamefont {{Miyatake}}}, \bibinfo {author} {\bibfnamefont
  {A.}~\bibnamefont {{Kannawadi}}}, \bibinfo {author} {\bibfnamefont
  {M.}~\bibnamefont {{Shirasaki}}}, \bibinfo {author} {\bibfnamefont
  {T.}~\bibnamefont {{Taniguchi}}}, \emph {et~al.},\ }\bibfield  {title}
  {\bibinfo {title} {{Hyper Suprime-Cam Year 3 results: Cosmology from cosmic
  shear power spectra}},\ }\href {https://doi.org/10.1103/PhysRevD.108.123519}
  {\bibfield  {journal} {\bibinfo  {journal} {\prd}\ }\textbf {\bibinfo
  {volume} {108}},\ \bibinfo {eid} {123519} (\bibinfo {year} {2023})},\ \Eprint
  {https://arxiv.org/abs/2304.00701} {arXiv:2304.00701 [astro-ph.CO]}
  \BibitemShut {NoStop}%
\bibitem [{\citenamefont {{Verde}}\ \emph {et~al.}(2019)\citenamefont
  {{Verde}}, \citenamefont {{Treu}},\ and\ \citenamefont
  {{Riess}}}]{Verde2019}%
  \BibitemOpen
  \bibfield  {author} {\bibinfo {author} {\bibfnamefont {L.}~\bibnamefont
  {{Verde}}}, \bibinfo {author} {\bibfnamefont {T.}~\bibnamefont {{Treu}}},\
  and\ \bibinfo {author} {\bibfnamefont {A.~G.}\ \bibnamefont {{Riess}}},\
  }\bibfield  {title} {\bibinfo {title} {{Tensions between the early and late
  Universe}},\ }\href {https://doi.org/10.1038/s41550-019-0902-0} {\bibfield
  {journal} {\bibinfo  {journal} {Nature Astronomy}\ }\textbf {\bibinfo
  {volume} {3}},\ \bibinfo {pages} {891} (\bibinfo {year} {2019})},\ \Eprint
  {https://arxiv.org/abs/1907.10625} {arXiv:1907.10625 [astro-ph.CO]}
  \BibitemShut {NoStop}%
\bibitem [{\citenamefont {{Knox}}\ and\ \citenamefont
  {{Millea}}(2020)}]{Knox2020}%
  \BibitemOpen
  \bibfield  {author} {\bibinfo {author} {\bibfnamefont {L.}~\bibnamefont
  {{Knox}}}\ and\ \bibinfo {author} {\bibfnamefont {M.}~\bibnamefont
  {{Millea}}},\ }\bibfield  {title} {\bibinfo {title} {{Hubble constant
  hunter's guide}},\ }\href {https://doi.org/10.1103/PhysRevD.101.043533}
  {\bibfield  {journal} {\bibinfo  {journal} {\prd}\ }\textbf {\bibinfo
  {volume} {101}},\ \bibinfo {eid} {043533} (\bibinfo {year} {2020})},\ \Eprint
  {https://arxiv.org/abs/1908.03663} {arXiv:1908.03663 [astro-ph.CO]}
  \BibitemShut {NoStop}%
\bibitem [{\citenamefont {{Di Valentino}}\ \emph
  {et~al.}(2021{\natexlab{a}})\citenamefont {{Di Valentino}}, \citenamefont
  {{Anchordoqui}}, \citenamefont {{Akarsu}}, \citenamefont {{Ali-Haimoud}},
  \citenamefont {{Amendola}}, \citenamefont {{Arendse}}, \citenamefont
  {{Asgari}}, \citenamefont {{Ballardini}}, \citenamefont {{Basilakos}},
  \citenamefont {{Battistelli}}, \citenamefont {{Benetti}}, \citenamefont
  {{Birrer}}, \citenamefont {{Bouchet}}, \citenamefont {{Bruni}}, \citenamefont
  {{Calabrese}}, \citenamefont {{Camarena}} \emph {et~al.}}]{Eleonora2021}%
  \BibitemOpen
  \bibfield  {author} {\bibinfo {author} {\bibfnamefont {E.}~\bibnamefont {{Di
  Valentino}}}, \bibinfo {author} {\bibfnamefont {L.~A.}\ \bibnamefont
  {{Anchordoqui}}}, \bibinfo {author} {\bibfnamefont {{\"O}.}~\bibnamefont
  {{Akarsu}}}, \bibinfo {author} {\bibfnamefont {Y.}~\bibnamefont
  {{Ali-Haimoud}}}, \bibinfo {author} {\bibfnamefont {L.}~\bibnamefont
  {{Amendola}}}, \bibinfo {author} {\bibfnamefont {N.}~\bibnamefont
  {{Arendse}}}, \bibinfo {author} {\bibfnamefont {M.}~\bibnamefont {{Asgari}}},
  \bibinfo {author} {\bibfnamefont {M.}~\bibnamefont {{Ballardini}}}, \bibinfo
  {author} {\bibfnamefont {S.}~\bibnamefont {{Basilakos}}}, \bibinfo {author}
  {\bibfnamefont {E.}~\bibnamefont {{Battistelli}}}, \bibinfo {author}
  {\bibfnamefont {M.}~\bibnamefont {{Benetti}}}, \bibinfo {author}
  {\bibfnamefont {S.}~\bibnamefont {{Birrer}}}, \bibinfo {author}
  {\bibfnamefont {F.~R.}\ \bibnamefont {{Bouchet}}}, \bibinfo {author}
  {\bibfnamefont {M.}~\bibnamefont {{Bruni}}}, \bibinfo {author} {\bibfnamefont
  {E.}~\bibnamefont {{Calabrese}}}, \bibinfo {author} {\bibfnamefont
  {D.}~\bibnamefont {{Camarena}}}, \emph {et~al.},\ }\bibfield  {title}
  {\bibinfo {title} {{Cosmology Intertwined II: The hubble constant tension}},\
  }\href {https://doi.org/10.1016/j.astropartphys.2021.102605} {\bibfield
  {journal} {\bibinfo  {journal} {Astroparticle Physics}\ }\textbf {\bibinfo
  {volume} {131}},\ \bibinfo {eid} {102605} (\bibinfo {year}
  {2021}{\natexlab{a}})},\ \Eprint {https://arxiv.org/abs/2008.11284}
  {arXiv:2008.11284 [astro-ph.CO]} \BibitemShut {NoStop}%
\bibitem [{\citenamefont {{Jedamzik}}\ \emph {et~al.}(2021)\citenamefont
  {{Jedamzik}}, \citenamefont {{Pogosian}},\ and\ \citenamefont
  {{Zhao}}}]{Jedamzik2021}%
  \BibitemOpen
  \bibfield  {author} {\bibinfo {author} {\bibfnamefont {K.}~\bibnamefont
  {{Jedamzik}}}, \bibinfo {author} {\bibfnamefont {L.}~\bibnamefont
  {{Pogosian}}},\ and\ \bibinfo {author} {\bibfnamefont {G.-B.}\ \bibnamefont
  {{Zhao}}},\ }\bibfield  {title} {\bibinfo {title} {{Why reducing the cosmic
  sound horizon alone can not fully resolve the Hubble tension}},\ }\href
  {https://doi.org/10.1038/s42005-021-00628-x} {\bibfield  {journal} {\bibinfo
  {journal} {Communications Physics}\ }\textbf {\bibinfo {volume} {4}},\
  \bibinfo {eid} {123} (\bibinfo {year} {2021})},\ \Eprint
  {https://arxiv.org/abs/2010.04158} {arXiv:2010.04158 [astro-ph.CO]}
  \BibitemShut {NoStop}%
\bibitem [{\citenamefont {{Di Valentino}}\ \emph
  {et~al.}(2021{\natexlab{b}})\citenamefont {{Di Valentino}}, \citenamefont
  {{Mena}}, \citenamefont {{Pan}}, \citenamefont {{Visinelli}}, \citenamefont
  {{Yang}}, \citenamefont {{Melchiorri}}, \citenamefont {{Mota}}, \citenamefont
  {{Riess}},\ and\ \citenamefont {{Silk}}}]{Eleonorab2021}%
  \BibitemOpen
  \bibfield  {author} {\bibinfo {author} {\bibfnamefont {E.}~\bibnamefont {{Di
  Valentino}}}, \bibinfo {author} {\bibfnamefont {O.}~\bibnamefont {{Mena}}},
  \bibinfo {author} {\bibfnamefont {S.}~\bibnamefont {{Pan}}}, \bibinfo
  {author} {\bibfnamefont {L.}~\bibnamefont {{Visinelli}}}, \bibinfo {author}
  {\bibfnamefont {W.}~\bibnamefont {{Yang}}}, \bibinfo {author} {\bibfnamefont
  {A.}~\bibnamefont {{Melchiorri}}}, \bibinfo {author} {\bibfnamefont {D.~F.}\
  \bibnamefont {{Mota}}}, \bibinfo {author} {\bibfnamefont {A.~G.}\
  \bibnamefont {{Riess}}},\ and\ \bibinfo {author} {\bibfnamefont
  {J.}~\bibnamefont {{Silk}}},\ }\bibfield  {title} {\bibinfo {title} {{In the
  realm of the Hubble tension-a review of solutions}},\ }\href
  {https://doi.org/10.1088/1361-6382/ac086d} {\bibfield  {journal} {\bibinfo
  {journal} {Classical and Quantum Gravity}\ }\textbf {\bibinfo {volume}
  {38}},\ \bibinfo {eid} {153001} (\bibinfo {year} {2021}{\natexlab{b}})},\
  \Eprint {https://arxiv.org/abs/2103.01183} {arXiv:2103.01183 [astro-ph.CO]}
  \BibitemShut {NoStop}%
\bibitem [{\citenamefont {{Perivolaropoulos}}\ and\ \citenamefont
  {{Skara}}(2022)}]{Perivolaropoulos2022}%
  \BibitemOpen
  \bibfield  {author} {\bibinfo {author} {\bibfnamefont {L.}~\bibnamefont
  {{Perivolaropoulos}}}\ and\ \bibinfo {author} {\bibfnamefont
  {F.}~\bibnamefont {{Skara}}},\ }\bibfield  {title} {\bibinfo {title}
  {{Challenges for {\ensuremath{\Lambda}}CDM: An update}},\ }\href
  {https://doi.org/10.1016/j.newar.2022.101659} {\bibfield  {journal} {\bibinfo
   {journal} {\nar}\ }\textbf {\bibinfo {volume} {95}},\ \bibinfo {eid}
  {101659} (\bibinfo {year} {2022})},\ \Eprint
  {https://arxiv.org/abs/2105.05208} {arXiv:2105.05208 [astro-ph.CO]}
  \BibitemShut {NoStop}%
\bibitem [{\citenamefont {{Shah}}\ \emph {et~al.}(2021)\citenamefont {{Shah}},
  \citenamefont {{Lemos}},\ and\ \citenamefont {{Lahav}}}]{Shah2021}%
  \BibitemOpen
  \bibfield  {author} {\bibinfo {author} {\bibfnamefont {P.}~\bibnamefont
  {{Shah}}}, \bibinfo {author} {\bibfnamefont {P.}~\bibnamefont {{Lemos}}},\
  and\ \bibinfo {author} {\bibfnamefont {O.}~\bibnamefont {{Lahav}}},\
  }\bibfield  {title} {\bibinfo {title} {{A buyer's guide to the Hubble
  constant}},\ }\href {https://doi.org/10.1007/s00159-021-00137-4} {\bibfield
  {journal} {\bibinfo  {journal} {\aapr}\ }\textbf {\bibinfo {volume} {29}},\
  \bibinfo {eid} {9} (\bibinfo {year} {2021})},\ \Eprint
  {https://arxiv.org/abs/2109.01161} {arXiv:2109.01161 [astro-ph.CO]}
  \BibitemShut {NoStop}%
\bibitem [{\citenamefont {{Abdalla}}\ \emph {et~al.}(2022)\citenamefont
  {{Abdalla}}, \citenamefont {{Abell{\'a}n}}, \citenamefont {{Aboubrahim}},
  \citenamefont {{Agnello}}, \citenamefont {{Akarsu}}, \citenamefont
  {{Akrami}}, \citenamefont {{Alestas}}, \citenamefont {{Aloni}}, \citenamefont
  {{Amendola}}, \citenamefont {{Anchordoqui}}, \citenamefont {{Anderson}},
  \citenamefont {{Arendse}}, \citenamefont {{Asgari}}, \citenamefont
  {{Ballardini}}, \citenamefont {{Barger}}, \citenamefont {{Basilakos}} \emph
  {et~al.}}]{Abdalla2022}%
  \BibitemOpen
  \bibfield  {author} {\bibinfo {author} {\bibfnamefont {E.}~\bibnamefont
  {{Abdalla}}}, \bibinfo {author} {\bibfnamefont {G.~F.}\ \bibnamefont
  {{Abell{\'a}n}}}, \bibinfo {author} {\bibfnamefont {A.}~\bibnamefont
  {{Aboubrahim}}}, \bibinfo {author} {\bibfnamefont {A.}~\bibnamefont
  {{Agnello}}}, \bibinfo {author} {\bibfnamefont {{\"O}.}~\bibnamefont
  {{Akarsu}}}, \bibinfo {author} {\bibfnamefont {Y.}~\bibnamefont {{Akrami}}},
  \bibinfo {author} {\bibfnamefont {G.}~\bibnamefont {{Alestas}}}, \bibinfo
  {author} {\bibfnamefont {D.}~\bibnamefont {{Aloni}}}, \bibinfo {author}
  {\bibfnamefont {L.}~\bibnamefont {{Amendola}}}, \bibinfo {author}
  {\bibfnamefont {L.~A.}\ \bibnamefont {{Anchordoqui}}}, \bibinfo {author}
  {\bibfnamefont {R.~I.}\ \bibnamefont {{Anderson}}}, \bibinfo {author}
  {\bibfnamefont {N.}~\bibnamefont {{Arendse}}}, \bibinfo {author}
  {\bibfnamefont {M.}~\bibnamefont {{Asgari}}}, \bibinfo {author}
  {\bibfnamefont {M.}~\bibnamefont {{Ballardini}}}, \bibinfo {author}
  {\bibfnamefont {V.}~\bibnamefont {{Barger}}}, \bibinfo {author}
  {\bibfnamefont {S.}~\bibnamefont {{Basilakos}}}, \emph {et~al.},\ }\bibfield
  {title} {\bibinfo {title} {{Cosmology intertwined: A review of the particle
  physics, astrophysics, and cosmology associated with the cosmological
  tensions and anomalies}},\ }\href
  {https://doi.org/10.1016/j.jheap.2022.04.002} {\bibfield  {journal} {\bibinfo
   {journal} {Journal of High Energy Astrophysics}\ }\textbf {\bibinfo {volume}
  {34}},\ \bibinfo {pages} {49} (\bibinfo {year} {2022})},\ \Eprint
  {https://arxiv.org/abs/2203.06142} {arXiv:2203.06142 [astro-ph.CO]}
  \BibitemShut {NoStop}%
\bibitem [{\citenamefont {{Hu}}(1998)}]{Hu1998}%
  \BibitemOpen
  \bibfield  {author} {\bibinfo {author} {\bibfnamefont {W.}~\bibnamefont
  {{Hu}}},\ }\bibfield  {title} {\bibinfo {title} {{Structure Formation with
  Generalized Dark Matter}},\ }\href {https://doi.org/10.1086/306274}
  {\bibfield  {journal} {\bibinfo  {journal} {\apj}\ }\textbf {\bibinfo
  {volume} {506}},\ \bibinfo {pages} {485} (\bibinfo {year} {1998})},\ \Eprint
  {https://arxiv.org/abs/astro-ph/9801234} {arXiv:astro-ph/9801234 [astro-ph]}
  \BibitemShut {NoStop}%
\bibitem [{\citenamefont {{Xu}}\ and\ \citenamefont {{Chang}}(2013)}]{Xu2013}%
  \BibitemOpen
  \bibfield  {author} {\bibinfo {author} {\bibfnamefont {L.}~\bibnamefont
  {{Xu}}}\ and\ \bibinfo {author} {\bibfnamefont {Y.}~\bibnamefont {{Chang}}},\
  }\bibfield  {title} {\bibinfo {title} {{Equation of state of dark matter
  after Planck data}},\ }\href {https://doi.org/10.1103/PhysRevD.88.127301}
  {\bibfield  {journal} {\bibinfo  {journal} {\prd}\ }\textbf {\bibinfo
  {volume} {88}},\ \bibinfo {eid} {127301} (\bibinfo {year} {2013})},\ \Eprint
  {https://arxiv.org/abs/1310.1532} {arXiv:1310.1532 [astro-ph.CO]}
  \BibitemShut {NoStop}%
\bibitem [{\citenamefont {{Kopp}}\ \emph {et~al.}(2018)\citenamefont {{Kopp}},
  \citenamefont {{Skordis}}, \citenamefont {{Thomas}},\ and\ \citenamefont
  {{Ili{\'c}}}}]{Kopp2018}%
  \BibitemOpen
  \bibfield  {author} {\bibinfo {author} {\bibfnamefont {M.}~\bibnamefont
  {{Kopp}}}, \bibinfo {author} {\bibfnamefont {C.}~\bibnamefont {{Skordis}}},
  \bibinfo {author} {\bibfnamefont {D.~B.}\ \bibnamefont {{Thomas}}},\ and\
  \bibinfo {author} {\bibfnamefont {S.}~\bibnamefont {{Ili{\'c}}}},\ }\bibfield
   {title} {\bibinfo {title} {{Dark Matter Equation of State through Cosmic
  History}},\ }\href {https://doi.org/10.1103/PhysRevLett.120.221102}
  {\bibfield  {journal} {\bibinfo  {journal} {\prl}\ }\textbf {\bibinfo
  {volume} {120}},\ \bibinfo {eid} {221102} (\bibinfo {year} {2018})},\ \Eprint
  {https://arxiv.org/abs/1802.09541} {arXiv:1802.09541 [astro-ph.CO]}
  \BibitemShut {NoStop}%
\bibitem [{\citenamefont {{Sachs}}\ and\ \citenamefont
  {{Wolfe}}(1967)}]{Sachswolfe1967}%
  \BibitemOpen
  \bibfield  {author} {\bibinfo {author} {\bibfnamefont {R.~K.}\ \bibnamefont
  {{Sachs}}}\ and\ \bibinfo {author} {\bibfnamefont {A.~M.}\ \bibnamefont
  {{Wolfe}}},\ }\bibfield  {title} {\bibinfo {title} {{Perturbations of a
  Cosmological Model and Angular Variations of the Microwave Background}},\
  }\href {https://doi.org/10.1086/148982} {\bibfield  {journal} {\bibinfo
  {journal} {\apj}\ }\textbf {\bibinfo {volume} {147}},\ \bibinfo {pages} {73}
  (\bibinfo {year} {1967})}\BibitemShut {NoStop}%
\bibitem [{\citenamefont {{Granett}}\ \emph {et~al.}(2008)\citenamefont
  {{Granett}}, \citenamefont {{Neyrinck}},\ and\ \citenamefont
  {{Szapudi}}}]{Granett2008}%
  \BibitemOpen
  \bibfield  {author} {\bibinfo {author} {\bibfnamefont {B.~R.}\ \bibnamefont
  {{Granett}}}, \bibinfo {author} {\bibfnamefont {M.~C.}\ \bibnamefont
  {{Neyrinck}}},\ and\ \bibinfo {author} {\bibfnamefont {I.}~\bibnamefont
  {{Szapudi}}},\ }\bibfield  {title} {\bibinfo {title} {{An Imprint of
  Superstructures on the Microwave Background due to the Integrated Sachs-Wolfe
  Effect}},\ }\href {https://doi.org/10.1086/591670} {\bibfield  {journal}
  {\bibinfo  {journal} {\apjl}\ }\textbf {\bibinfo {volume} {683}},\ \bibinfo
  {pages} {L99} (\bibinfo {year} {2008})},\ \Eprint
  {https://arxiv.org/abs/0805.3695} {arXiv:0805.3695 [astro-ph]} \BibitemShut
  {NoStop}%
\bibitem [{\citenamefont {{Cai}}\ \emph {et~al.}(2014)\citenamefont {{Cai}},
  \citenamefont {{Neyrinck}}, \citenamefont {{Szapudi}}, \citenamefont
  {{Cole}},\ and\ \citenamefont {{Frenk}}}]{Cai2014}%
  \BibitemOpen
  \bibfield  {author} {\bibinfo {author} {\bibfnamefont {Y.-C.}\ \bibnamefont
  {{Cai}}}, \bibinfo {author} {\bibfnamefont {M.~C.}\ \bibnamefont
  {{Neyrinck}}}, \bibinfo {author} {\bibfnamefont {I.}~\bibnamefont
  {{Szapudi}}}, \bibinfo {author} {\bibfnamefont {S.}~\bibnamefont {{Cole}}},\
  and\ \bibinfo {author} {\bibfnamefont {C.~S.}\ \bibnamefont {{Frenk}}},\
  }\bibfield  {title} {\bibinfo {title} {{A Possible Cold Imprint of Voids on
  the Microwave Background Radiation}},\ }\href
  {https://doi.org/10.1088/0004-637X/786/2/110} {\bibfield  {journal} {\bibinfo
   {journal} {\apj}\ }\textbf {\bibinfo {volume} {786}},\ \bibinfo {eid} {110}
  (\bibinfo {year} {2014})},\ \Eprint {https://arxiv.org/abs/1301.6136}
  {arXiv:1301.6136 [astro-ph.CO]} \BibitemShut {NoStop}%
\bibitem [{\citenamefont {{Kov{\'a}cs}}(2018)}]{Kovacs2018}%
  \BibitemOpen
  \bibfield  {author} {\bibinfo {author} {\bibfnamefont {A.}~\bibnamefont
  {{Kov{\'a}cs}}},\ }\bibfield  {title} {\bibinfo {title} {{The part and the
  whole: voids, supervoids, and their ISW imprint}},\ }\href
  {https://doi.org/10.1093/mnras/stx3213} {\bibfield  {journal} {\bibinfo
  {journal} {\mnras}\ }\textbf {\bibinfo {volume} {475}},\ \bibinfo {pages}
  {1777} (\bibinfo {year} {2018})},\ \Eprint {https://arxiv.org/abs/1701.08583}
  {arXiv:1701.08583 [astro-ph.CO]} \BibitemShut {NoStop}%
\bibitem [{\citenamefont {{Kov{\'a}cs}}\ \emph {et~al.}(2019)\citenamefont
  {{Kov{\'a}cs}}, \citenamefont {{S{\'a}nchez}}, \citenamefont
  {{Garc{\'\i}a-Bellido}}, \citenamefont {{Elvin-Poole}}, \citenamefont
  {{Hamaus}}, \citenamefont {{Miranda}}, \citenamefont {{Nadathur}},
  \citenamefont {{Abbott}}, \citenamefont {{Abdalla}}, \citenamefont {{Annis}},
  \citenamefont {{Avila}}, \citenamefont {{Bertin}}, \citenamefont {{Brooks}},
  \citenamefont {{Burke}}, \citenamefont {{Carnero Rosell}} \emph
  {et~al.}}]{Kovacs2019}%
  \BibitemOpen
  \bibfield  {author} {\bibinfo {author} {\bibfnamefont {A.}~\bibnamefont
  {{Kov{\'a}cs}}}, \bibinfo {author} {\bibfnamefont {C.}~\bibnamefont
  {{S{\'a}nchez}}}, \bibinfo {author} {\bibfnamefont {J.}~\bibnamefont
  {{Garc{\'\i}a-Bellido}}}, \bibinfo {author} {\bibfnamefont {J.}~\bibnamefont
  {{Elvin-Poole}}}, \bibinfo {author} {\bibfnamefont {N.}~\bibnamefont
  {{Hamaus}}}, \bibinfo {author} {\bibfnamefont {V.}~\bibnamefont {{Miranda}}},
  \bibinfo {author} {\bibfnamefont {S.}~\bibnamefont {{Nadathur}}}, \bibinfo
  {author} {\bibfnamefont {T.}~\bibnamefont {{Abbott}}}, \bibinfo {author}
  {\bibfnamefont {F.~B.}\ \bibnamefont {{Abdalla}}}, \bibinfo {author}
  {\bibfnamefont {J.}~\bibnamefont {{Annis}}}, \bibinfo {author} {\bibfnamefont
  {S.}~\bibnamefont {{Avila}}}, \bibinfo {author} {\bibfnamefont
  {E.}~\bibnamefont {{Bertin}}}, \bibinfo {author} {\bibfnamefont
  {D.}~\bibnamefont {{Brooks}}}, \bibinfo {author} {\bibfnamefont {D.~L.}\
  \bibnamefont {{Burke}}}, \bibinfo {author} {\bibfnamefont {A.}~\bibnamefont
  {{Carnero Rosell}}}, \emph {et~al.},\ }\bibfield  {title} {\bibinfo {title}
  {{More out of less: an excess integrated Sachs-Wolfe signal from supervoids
  mapped out by the Dark Energy Survey}},\ }\href
  {https://doi.org/10.1093/mnras/stz341} {\bibfield  {journal} {\bibinfo
  {journal} {\mnras}\ }\textbf {\bibinfo {volume} {484}},\ \bibinfo {pages}
  {5267} (\bibinfo {year} {2019})},\ \Eprint {https://arxiv.org/abs/1811.07812}
  {arXiv:1811.07812 [astro-ph.CO]} \BibitemShut {NoStop}%
\bibitem [{\citenamefont {{Nadathur}}\ and\ \citenamefont
  {{Crittenden}}(2016)}]{specz_params3}%
  \BibitemOpen
  \bibfield  {author} {\bibinfo {author} {\bibfnamefont {S.}~\bibnamefont
  {{Nadathur}}}\ and\ \bibinfo {author} {\bibfnamefont {R.}~\bibnamefont
  {{Crittenden}}},\ }\bibfield  {title} {\bibinfo {title} {{A Detection of the
  Integrated Sachs-Wolfe Imprint of Cosmic Superstructures Using a
  Matched-filter Approach}},\ }\href
  {https://doi.org/10.3847/2041-8205/830/1/L19} {\bibfield  {journal} {\bibinfo
   {journal} {\apjl}\ }\textbf {\bibinfo {volume} {830}},\ \bibinfo {eid} {L19}
  (\bibinfo {year} {2016})},\ \Eprint {https://arxiv.org/abs/1608.08638}
  {arXiv:1608.08638 [astro-ph.CO]} \BibitemShut {NoStop}%
\bibitem [{\citenamefont {{Kov{\'a}cs}}\ \emph
  {et~al.}(2022{\natexlab{a}})\citenamefont {{Kov{\'a}cs}}, \citenamefont
  {{Vielzeuf}}, \citenamefont {{Ferrero}}, \citenamefont {{Fosalba}},
  \citenamefont {{Demirbozan}}, \citenamefont {{Miquel}}, \citenamefont
  {{Chang}}, \citenamefont {{Hamaus}}, \citenamefont {{Pollina}}, \citenamefont
  {{Bechtol}}, \citenamefont {{Becker}}, \citenamefont {{Carnero Rosell}},
  \citenamefont {{Carrasco Kind}}, \citenamefont {{Cawthon}}, \citenamefont
  {{Crocce}} \emph {et~al.}}]{Kovacs2022}%
  \BibitemOpen
  \bibfield  {author} {\bibinfo {author} {\bibfnamefont {A.}~\bibnamefont
  {{Kov{\'a}cs}}}, \bibinfo {author} {\bibfnamefont {P.}~\bibnamefont
  {{Vielzeuf}}}, \bibinfo {author} {\bibfnamefont {I.}~\bibnamefont
  {{Ferrero}}}, \bibinfo {author} {\bibfnamefont {P.}~\bibnamefont
  {{Fosalba}}}, \bibinfo {author} {\bibfnamefont {U.}~\bibnamefont
  {{Demirbozan}}}, \bibinfo {author} {\bibfnamefont {R.}~\bibnamefont
  {{Miquel}}}, \bibinfo {author} {\bibfnamefont {C.}~\bibnamefont {{Chang}}},
  \bibinfo {author} {\bibfnamefont {N.}~\bibnamefont {{Hamaus}}}, \bibinfo
  {author} {\bibfnamefont {G.}~\bibnamefont {{Pollina}}}, \bibinfo {author}
  {\bibfnamefont {K.}~\bibnamefont {{Bechtol}}}, \bibinfo {author}
  {\bibfnamefont {M.}~\bibnamefont {{Becker}}}, \bibinfo {author}
  {\bibfnamefont {A.}~\bibnamefont {{Carnero Rosell}}}, \bibinfo {author}
  {\bibfnamefont {M.}~\bibnamefont {{Carrasco Kind}}}, \bibinfo {author}
  {\bibfnamefont {R.}~\bibnamefont {{Cawthon}}}, \bibinfo {author}
  {\bibfnamefont {M.}~\bibnamefont {{Crocce}}}, \emph {et~al.},\ }\bibfield
  {title} {\bibinfo {title} {{Dark Energy Survey Year 3 results: imprints of
  cosmic voids and superclusters in the Planck CMB lensing map}},\ }\href@noop
  {} {\bibfield  {journal} {\bibinfo  {journal} {arXiv e-prints}\ ,\ \bibinfo
  {eid} {arXiv:2203.11306}} (\bibinfo {year} {2022}{\natexlab{a}})},\ \Eprint
  {https://arxiv.org/abs/2203.11306} {arXiv:2203.11306 [astro-ph.CO]}
  \BibitemShut {NoStop}%
\bibitem [{\citenamefont {{de Cesare}}\ and\ \citenamefont
  {{Wilson-Ewing}}(2022)}]{Cesare2022}%
  \BibitemOpen
  \bibfield  {author} {\bibinfo {author} {\bibfnamefont {M.}~\bibnamefont {{de
  Cesare}}}\ and\ \bibinfo {author} {\bibfnamefont {E.}~\bibnamefont
  {{Wilson-Ewing}}},\ }\bibfield  {title} {\bibinfo {title} {{Interacting dark
  sector from the trace-free Einstein equations: Cosmological perturbations
  with no instability}},\ }\href {https://doi.org/10.1103/PhysRevD.106.023527}
  {\bibfield  {journal} {\bibinfo  {journal} {\prd}\ }\textbf {\bibinfo
  {volume} {106}},\ \bibinfo {eid} {023527} (\bibinfo {year} {2022})},\ \Eprint
  {https://arxiv.org/abs/2112.12701} {arXiv:2112.12701 [gr-qc]} \BibitemShut
  {NoStop}%
\bibitem [{\citenamefont {{Farrar}}\ and\ \citenamefont
  {{Peebles}}(2004)}]{Farrar2004}%
  \BibitemOpen
  \bibfield  {author} {\bibinfo {author} {\bibfnamefont {G.~R.}\ \bibnamefont
  {{Farrar}}}\ and\ \bibinfo {author} {\bibfnamefont {P.~J.~E.}\ \bibnamefont
  {{Peebles}}},\ }\bibfield  {title} {\bibinfo {title} {{Interacting Dark
  Matter and Dark Energy}},\ }\href {https://doi.org/10.1086/381728} {\bibfield
   {journal} {\bibinfo  {journal} {\apj}\ }\textbf {\bibinfo {volume} {604}},\
  \bibinfo {pages} {1} (\bibinfo {year} {2004})},\ \Eprint
  {https://arxiv.org/abs/astro-ph/0307316} {arXiv:astro-ph/0307316 [astro-ph]}
  \BibitemShut {NoStop}%
\bibitem [{\citenamefont {{Peebles}}(2012)}]{Peebles2012}%
  \BibitemOpen
  \bibfield  {author} {\bibinfo {author} {\bibfnamefont {P.~J.~E.}\
  \bibnamefont {{Peebles}}},\ }\bibfield  {title} {\bibinfo {title}
  {{Evanescent matter}},\ }\href {https://doi.org/10.1002/andp.201200072}
  {\bibfield  {journal} {\bibinfo  {journal} {Annalen der Physik}\ }\textbf
  {\bibinfo {volume} {524}},\ \bibinfo {pages} {591} (\bibinfo {year}
  {2012})},\ \Eprint {https://arxiv.org/abs/1204.0485} {arXiv:1204.0485
  [astro-ph.CO]} \BibitemShut {NoStop}%
\bibitem [{\citenamefont {{Blas}}\ \emph {et~al.}(2011)\citenamefont {{Blas}},
  \citenamefont {{Lesgourgues}},\ and\ \citenamefont {{Tram}}}]{CLASS}%
  \BibitemOpen
  \bibfield  {author} {\bibinfo {author} {\bibfnamefont {D.}~\bibnamefont
  {{Blas}}}, \bibinfo {author} {\bibfnamefont {J.}~\bibnamefont
  {{Lesgourgues}}},\ and\ \bibinfo {author} {\bibfnamefont {T.}~\bibnamefont
  {{Tram}}},\ }\bibfield  {title} {\bibinfo {title} {{The Cosmic Linear
  Anisotropy Solving System (CLASS). Part II: Approximation schemes}},\ }\href
  {https://doi.org/10.1088/1475-7516/2011/07/034} {\bibfield  {journal}
  {\bibinfo  {journal} {\jcap}\ }\textbf {\bibinfo {volume} {2011}},\ \bibinfo
  {eid} {034} (\bibinfo {year} {2011})},\ \Eprint
  {https://arxiv.org/abs/1104.2933} {arXiv:1104.2933 [astro-ph.CO]}
  \BibitemShut {NoStop}%
\bibitem [{\citenamefont {{Riess}}\ \emph
  {et~al.}(2021{\natexlab{a}})\citenamefont {{Riess}}, \citenamefont {{Yuan}},
  \citenamefont {{Macri}}, \citenamefont {{Scolnic}}, \citenamefont {{Brout}},
  \citenamefont {{Casertano}}, \citenamefont {{Jones}}, \citenamefont
  {{Murakami}}, \citenamefont {{Breuval}}, \citenamefont {{Brink}},
  \citenamefont {{Filippenko}}, \citenamefont {{Hoffmann}}, \citenamefont
  {{Jha}}, \citenamefont {{Kenworthy}}, \citenamefont {{Mackenty}} \emph
  {et~al.}}]{Riess2021}%
  \BibitemOpen
  \bibfield  {author} {\bibinfo {author} {\bibfnamefont {A.~G.}\ \bibnamefont
  {{Riess}}}, \bibinfo {author} {\bibfnamefont {W.}~\bibnamefont {{Yuan}}},
  \bibinfo {author} {\bibfnamefont {L.~M.}\ \bibnamefont {{Macri}}}, \bibinfo
  {author} {\bibfnamefont {D.}~\bibnamefont {{Scolnic}}}, \bibinfo {author}
  {\bibfnamefont {D.}~\bibnamefont {{Brout}}}, \bibinfo {author} {\bibfnamefont
  {S.}~\bibnamefont {{Casertano}}}, \bibinfo {author} {\bibfnamefont {D.~O.}\
  \bibnamefont {{Jones}}}, \bibinfo {author} {\bibfnamefont {Y.}~\bibnamefont
  {{Murakami}}}, \bibinfo {author} {\bibfnamefont {L.}~\bibnamefont
  {{Breuval}}}, \bibinfo {author} {\bibfnamefont {T.~G.}\ \bibnamefont
  {{Brink}}}, \bibinfo {author} {\bibfnamefont {A.~V.}\ \bibnamefont
  {{Filippenko}}}, \bibinfo {author} {\bibfnamefont {S.}~\bibnamefont
  {{Hoffmann}}}, \bibinfo {author} {\bibfnamefont {S.~W.}\ \bibnamefont
  {{Jha}}}, \bibinfo {author} {\bibfnamefont {W.~D.}\ \bibnamefont
  {{Kenworthy}}}, \bibinfo {author} {\bibfnamefont {J.}~\bibnamefont
  {{Mackenty}}}, \emph {et~al.},\ }\bibfield  {title} {\bibinfo {title} {{A
  Comprehensive Measurement of the Local Value of the Hubble Constant with 1
  km/s/Mpc Uncertainty from the Hubble Space Telescope and the SH0ES Team}},\
  }\href@noop {} {\bibfield  {journal} {\bibinfo  {journal} {arXiv e-prints}\
  ,\ \bibinfo {eid} {arXiv:2112.04510}} (\bibinfo {year}
  {2021}{\natexlab{a}})},\ \Eprint {https://arxiv.org/abs/2112.04510}
  {arXiv:2112.04510 [astro-ph.CO]} \BibitemShut {NoStop}%
\bibitem [{\citenamefont {{Hang}}\ \emph {et~al.}(2021)\citenamefont {{Hang}},
  \citenamefont {{Alam}}, \citenamefont {{Peacock}},\ and\ \citenamefont
  {{Cai}}}]{Hang2021}%
  \BibitemOpen
  \bibfield  {author} {\bibinfo {author} {\bibfnamefont {Q.}~\bibnamefont
  {{Hang}}}, \bibinfo {author} {\bibfnamefont {S.}~\bibnamefont {{Alam}}},
  \bibinfo {author} {\bibfnamefont {J.~A.}\ \bibnamefont {{Peacock}}},\ and\
  \bibinfo {author} {\bibfnamefont {Y.-C.}\ \bibnamefont {{Cai}}},\ }\bibfield
  {title} {\bibinfo {title} {{Galaxy clustering in the DESI Legacy Survey and
  its imprint on the CMB}},\ }\href {https://doi.org/10.1093/mnras/staa3738}
  {\bibfield  {journal} {\bibinfo  {journal} {\mnras}\ }\textbf {\bibinfo
  {volume} {501}},\ \bibinfo {pages} {1481} (\bibinfo {year} {2021})},\ \Eprint
  {https://arxiv.org/abs/2010.00466} {arXiv:2010.00466 [astro-ph.CO]}
  \BibitemShut {NoStop}%
\bibitem [{\citenamefont {{Kov{\'a}cs}}\ and\ \citenamefont
  {{Garc{\'\i}a-Bellido}}(2016)}]{EllipticalVoid}%
  \BibitemOpen
  \bibfield  {author} {\bibinfo {author} {\bibfnamefont {A.}~\bibnamefont
  {{Kov{\'a}cs}}}\ and\ \bibinfo {author} {\bibfnamefont {J.}~\bibnamefont
  {{Garc{\'\i}a-Bellido}}},\ }\bibfield  {title} {\bibinfo {title} {{Cosmic
  troublemakers: the Cold Spot, the Eridanus supervoid, and the Great Walls}},\
  }\href {https://doi.org/10.1093/mnras/stw1752} {\bibfield  {journal}
  {\bibinfo  {journal} {\mnras}\ }\textbf {\bibinfo {volume} {462}},\ \bibinfo
  {pages} {1882} (\bibinfo {year} {2016})},\ \Eprint
  {https://arxiv.org/abs/1511.09008} {arXiv:1511.09008 [astro-ph.CO]}
  \BibitemShut {NoStop}%
\bibitem [{\citenamefont {{Nadathur}}\ and\ \citenamefont
  {{Hotchkiss}}(2015)}]{specz_params1}%
  \BibitemOpen
  \bibfield  {author} {\bibinfo {author} {\bibfnamefont {S.}~\bibnamefont
  {{Nadathur}}}\ and\ \bibinfo {author} {\bibfnamefont {S.}~\bibnamefont
  {{Hotchkiss}}},\ }\bibfield  {title} {\bibinfo {title} {{The nature of voids
  - II. Tracing underdensities with biased galaxies}},\ }\href
  {https://doi.org/10.1093/mnras/stv1994} {\bibfield  {journal} {\bibinfo
  {journal} {\mnras}\ }\textbf {\bibinfo {volume} {454}},\ \bibinfo {pages}
  {889} (\bibinfo {year} {2015})},\ \Eprint {https://arxiv.org/abs/1507.00197}
  {arXiv:1507.00197 [astro-ph.CO]} \BibitemShut {NoStop}%
\bibitem [{\citenamefont {{Nadathur}}(2016)}]{specz_params2}%
  \BibitemOpen
  \bibfield  {author} {\bibinfo {author} {\bibfnamefont {S.}~\bibnamefont
  {{Nadathur}}},\ }\bibfield  {title} {\bibinfo {title} {{Testing cosmology
  with a catalogue of voids in the BOSS galaxy surveys}},\ }\href
  {https://doi.org/10.1093/mnras/stw1340} {\bibfield  {journal} {\bibinfo
  {journal} {\mnras}\ }\textbf {\bibinfo {volume} {461}},\ \bibinfo {pages}
  {358} (\bibinfo {year} {2016})},\ \Eprint {https://arxiv.org/abs/1602.04752}
  {arXiv:1602.04752 [astro-ph.CO]} \BibitemShut {NoStop}%
\bibitem [{\citenamefont {{Raghunathan}}\ \emph {et~al.}(2020)\citenamefont
  {{Raghunathan}}, \citenamefont {{Nadathur}}, \citenamefont {{Sherwin}},\ and\
  \citenamefont {{Whitehorn}}}]{specz_params4}%
  \BibitemOpen
  \bibfield  {author} {\bibinfo {author} {\bibfnamefont {S.}~\bibnamefont
  {{Raghunathan}}}, \bibinfo {author} {\bibfnamefont {S.}~\bibnamefont
  {{Nadathur}}}, \bibinfo {author} {\bibfnamefont {B.~D.}\ \bibnamefont
  {{Sherwin}}},\ and\ \bibinfo {author} {\bibfnamefont {N.}~\bibnamefont
  {{Whitehorn}}},\ }\bibfield  {title} {\bibinfo {title} {{The Gravitational
  Lensing Signatures of BOSS Voids in the Cosmic Microwave Background}},\
  }\href {https://doi.org/10.3847/1538-4357/ab6f05} {\bibfield  {journal}
  {\bibinfo  {journal} {\apj}\ }\textbf {\bibinfo {volume} {890}},\ \bibinfo
  {eid} {168} (\bibinfo {year} {2020})},\ \Eprint
  {https://arxiv.org/abs/1911.08475} {arXiv:1911.08475 [astro-ph.CO]}
  \BibitemShut {NoStop}%
\bibitem [{\citenamefont {{Kov{\'a}cs}}\ \emph {et~al.}(2017)\citenamefont
  {{Kov{\'a}cs}}, \citenamefont {{S{\'a}nchez}}, \citenamefont
  {{Garc{\'\i}a-Bellido}}, \citenamefont {{Nadathur}}, \citenamefont
  {{Crittenden}}, \citenamefont {{Gruen}}, \citenamefont {{Huterer}},
  \citenamefont {{Bacon}}, \citenamefont {{Clampitt}}, \citenamefont
  {{DeRose}}, \citenamefont {{Dodelson}}, \citenamefont {{Gazta{\~n}aga}},
  \citenamefont {{Jain}}, \citenamefont {{Kirk}} \emph
  {et~al.}}]{photoz_params1}%
  \BibitemOpen
  \bibfield  {author} {\bibinfo {author} {\bibfnamefont {A.}~\bibnamefont
  {{Kov{\'a}cs}}}, \bibinfo {author} {\bibfnamefont {C.}~\bibnamefont
  {{S{\'a}nchez}}}, \bibinfo {author} {\bibfnamefont {J.}~\bibnamefont
  {{Garc{\'\i}a-Bellido}}}, \bibinfo {author} {\bibfnamefont {S.}~\bibnamefont
  {{Nadathur}}}, \bibinfo {author} {\bibfnamefont {R.}~\bibnamefont
  {{Crittenden}}}, \bibinfo {author} {\bibfnamefont {D.}~\bibnamefont
  {{Gruen}}}, \bibinfo {author} {\bibfnamefont {D.}~\bibnamefont {{Huterer}}},
  \bibinfo {author} {\bibfnamefont {D.}~\bibnamefont {{Bacon}}}, \bibinfo
  {author} {\bibfnamefont {J.}~\bibnamefont {{Clampitt}}}, \bibinfo {author}
  {\bibfnamefont {J.}~\bibnamefont {{DeRose}}}, \bibinfo {author}
  {\bibfnamefont {S.}~\bibnamefont {{Dodelson}}}, \bibinfo {author}
  {\bibfnamefont {E.}~\bibnamefont {{Gazta{\~n}aga}}}, \bibinfo {author}
  {\bibfnamefont {B.}~\bibnamefont {{Jain}}}, \bibinfo {author} {\bibfnamefont
  {D.}~\bibnamefont {{Kirk}}}, \emph {et~al.},\ }\bibfield  {title} {\bibinfo
  {title} {{Imprint of DES superstructures on the cosmic microwave
  background}},\ }\href {https://doi.org/10.1093/mnras/stw2968} {\bibfield
  {journal} {\bibinfo  {journal} {\mnras}\ }\textbf {\bibinfo {volume} {465}},\
  \bibinfo {pages} {4166} (\bibinfo {year} {2017})},\ \Eprint
  {https://arxiv.org/abs/1610.00637} {arXiv:1610.00637 [astro-ph.CO]}
  \BibitemShut {NoStop}%
\bibitem [{\citenamefont {{Mackenzie}}\ \emph {et~al.}(2017)\citenamefont
  {{Mackenzie}}, \citenamefont {{Shanks}}, \citenamefont {{Bremer}},
  \citenamefont {{Cai}}, \citenamefont {{Gunawardhana}}, \citenamefont
  {{Kov{\'a}cs}}, \citenamefont {{Norberg}},\ and\ \citenamefont
  {{Szapudi}}}]{Mackenzie2017}%
  \BibitemOpen
  \bibfield  {author} {\bibinfo {author} {\bibfnamefont {R.}~\bibnamefont
  {{Mackenzie}}}, \bibinfo {author} {\bibfnamefont {T.}~\bibnamefont
  {{Shanks}}}, \bibinfo {author} {\bibfnamefont {M.~N.}\ \bibnamefont
  {{Bremer}}}, \bibinfo {author} {\bibfnamefont {Y.-C.}\ \bibnamefont {{Cai}}},
  \bibinfo {author} {\bibfnamefont {M.~L.~P.}\ \bibnamefont {{Gunawardhana}}},
  \bibinfo {author} {\bibfnamefont {A.}~\bibnamefont {{Kov{\'a}cs}}}, \bibinfo
  {author} {\bibfnamefont {P.}~\bibnamefont {{Norberg}}},\ and\ \bibinfo
  {author} {\bibfnamefont {I.}~\bibnamefont {{Szapudi}}},\ }\bibfield  {title}
  {\bibinfo {title} {{Evidence against a supervoid causing the CMB Cold
  Spot}},\ }\href {https://doi.org/10.1093/mnras/stx931} {\bibfield  {journal}
  {\bibinfo  {journal} {\mnras}\ }\textbf {\bibinfo {volume} {470}},\ \bibinfo
  {pages} {2328} (\bibinfo {year} {2017})},\ \Eprint
  {https://arxiv.org/abs/1704.03814} {arXiv:1704.03814 [astro-ph.CO]}
  \BibitemShut {NoStop}%
\bibitem [{\citenamefont {{Vielva}}\ \emph {et~al.}(2004)\citenamefont
  {{Vielva}}, \citenamefont {{Mart{\'\i}nez-Gonz{\'a}lez}}, \citenamefont
  {{Barreiro}}, \citenamefont {{Sanz}},\ and\ \citenamefont
  {{Cay{\'o}n}}}]{Vielva2004}%
  \BibitemOpen
  \bibfield  {author} {\bibinfo {author} {\bibfnamefont {P.}~\bibnamefont
  {{Vielva}}}, \bibinfo {author} {\bibfnamefont {E.}~\bibnamefont
  {{Mart{\'\i}nez-Gonz{\'a}lez}}}, \bibinfo {author} {\bibfnamefont {R.~B.}\
  \bibnamefont {{Barreiro}}}, \bibinfo {author} {\bibfnamefont {J.~L.}\
  \bibnamefont {{Sanz}}},\ and\ \bibinfo {author} {\bibfnamefont
  {L.}~\bibnamefont {{Cay{\'o}n}}},\ }\bibfield  {title} {\bibinfo {title}
  {{Detection of Non-Gaussianity in the Wilkinson Microwave Anisotropy Probe
  First-Year Data Using Spherical Wavelets}},\ }\href
  {https://doi.org/10.1086/421007} {\bibfield  {journal} {\bibinfo  {journal}
  {\apj}\ }\textbf {\bibinfo {volume} {609}},\ \bibinfo {pages} {22} (\bibinfo
  {year} {2004})},\ \Eprint {https://arxiv.org/abs/astro-ph/0310273}
  {arXiv:astro-ph/0310273 [astro-ph]} \BibitemShut {NoStop}%
\bibitem [{\citenamefont {{Inoue}}\ and\ \citenamefont
  {{Silk}}(2006)}]{Inoue2006}%
  \BibitemOpen
  \bibfield  {author} {\bibinfo {author} {\bibfnamefont {K.~T.}\ \bibnamefont
  {{Inoue}}}\ and\ \bibinfo {author} {\bibfnamefont {J.}~\bibnamefont
  {{Silk}}},\ }\bibfield  {title} {\bibinfo {title} {{Local Voids as the Origin
  of Large-Angle Cosmic Microwave Background Anomalies. I.}},\ }\href
  {https://doi.org/10.1086/505636} {\bibfield  {journal} {\bibinfo  {journal}
  {\apj}\ }\textbf {\bibinfo {volume} {648}},\ \bibinfo {pages} {23} (\bibinfo
  {year} {2006})},\ \Eprint {https://arxiv.org/abs/astro-ph/0602478}
  {arXiv:astro-ph/0602478 [astro-ph]} \BibitemShut {NoStop}%
\bibitem [{\citenamefont {{Inoue}}\ and\ \citenamefont
  {{Silk}}(2007)}]{Inoue2007}%
  \BibitemOpen
  \bibfield  {author} {\bibinfo {author} {\bibfnamefont {K.~T.}\ \bibnamefont
  {{Inoue}}}\ and\ \bibinfo {author} {\bibfnamefont {J.}~\bibnamefont
  {{Silk}}},\ }\bibfield  {title} {\bibinfo {title} {{Local Voids as the Origin
  of Large-Angle Cosmic Microwave Background Anomalies: The Effect of a
  Cosmological Constant}},\ }\href {https://doi.org/10.1086/517603} {\bibfield
  {journal} {\bibinfo  {journal} {\apj}\ }\textbf {\bibinfo {volume} {664}},\
  \bibinfo {pages} {650} (\bibinfo {year} {2007})},\ \Eprint
  {https://arxiv.org/abs/astro-ph/0612347} {arXiv:astro-ph/0612347 [astro-ph]}
  \BibitemShut {NoStop}%
\bibitem [{\citenamefont {{Masina}}\ and\ \citenamefont
  {{Notari}}(2009)}]{Masina2009}%
  \BibitemOpen
  \bibfield  {author} {\bibinfo {author} {\bibfnamefont {I.}~\bibnamefont
  {{Masina}}}\ and\ \bibinfo {author} {\bibfnamefont {A.}~\bibnamefont
  {{Notari}}},\ }\bibfield  {title} {\bibinfo {title} {{The cold spot as a
  large void: Rees-Sciama effect on CMB power spectrum and bispectrum}},\
  }\href {https://doi.org/10.1088/1475-7516/2009/02/019} {\bibfield  {journal}
  {\bibinfo  {journal} {\jcap}\ }\textbf {\bibinfo {volume} {2009}},\ \bibinfo
  {eid} {019} (\bibinfo {year} {2009})},\ \Eprint
  {https://arxiv.org/abs/0808.1811} {arXiv:0808.1811 [astro-ph]} \BibitemShut
  {NoStop}%
\bibitem [{\citenamefont {{Nadathur}}\ \emph {et~al.}(2014)\citenamefont
  {{Nadathur}}, \citenamefont {{Lavinto}}, \citenamefont {{Hotchkiss}},\ and\
  \citenamefont {{R{\"a}s{\"a}nen}}}]{Nadathur2014}%
  \BibitemOpen
  \bibfield  {author} {\bibinfo {author} {\bibfnamefont {S.}~\bibnamefont
  {{Nadathur}}}, \bibinfo {author} {\bibfnamefont {M.}~\bibnamefont
  {{Lavinto}}}, \bibinfo {author} {\bibfnamefont {S.}~\bibnamefont
  {{Hotchkiss}}},\ and\ \bibinfo {author} {\bibfnamefont {S.}~\bibnamefont
  {{R{\"a}s{\"a}nen}}},\ }\bibfield  {title} {\bibinfo {title} {{Can a
  supervoid explain the cold spot?}},\ }\href
  {https://doi.org/10.1103/PhysRevD.90.103510} {\bibfield  {journal} {\bibinfo
  {journal} {\prd}\ }\textbf {\bibinfo {volume} {90}},\ \bibinfo {eid} {103510}
  (\bibinfo {year} {2014})},\ \Eprint {https://arxiv.org/abs/1408.4720}
  {arXiv:1408.4720 [astro-ph.CO]} \BibitemShut {NoStop}%
\bibitem [{\citenamefont {{Naidoo}}\ \emph {et~al.}(2016)\citenamefont
  {{Naidoo}}, \citenamefont {{Benoit-L{\'e}vy}},\ and\ \citenamefont
  {{Lahav}}}]{Naidoo2016}%
  \BibitemOpen
  \bibfield  {author} {\bibinfo {author} {\bibfnamefont {K.}~\bibnamefont
  {{Naidoo}}}, \bibinfo {author} {\bibfnamefont {A.}~\bibnamefont
  {{Benoit-L{\'e}vy}}},\ and\ \bibinfo {author} {\bibfnamefont
  {O.}~\bibnamefont {{Lahav}}},\ }\bibfield  {title} {\bibinfo {title} {{Could
  multiple voids explain the cosmic microwave background Cold Spot anomaly?}},\
  }\href {https://doi.org/10.1093/mnrasl/slw043} {\bibfield  {journal}
  {\bibinfo  {journal} {\mnras}\ }\textbf {\bibinfo {volume} {459}},\ \bibinfo
  {pages} {L71} (\bibinfo {year} {2016})},\ \Eprint
  {https://arxiv.org/abs/1512.02694} {arXiv:1512.02694 [astro-ph.CO]}
  \BibitemShut {NoStop}%
\bibitem [{\citenamefont {{Marcos-Caballero}}\ \emph
  {et~al.}(2016)\citenamefont {{Marcos-Caballero}}, \citenamefont
  {{Fern{\'a}ndez-Cobos}}, \citenamefont {{Mart{\'\i}nez-Gonz{\'a}lez}},\ and\
  \citenamefont {{Vielva}}}]{Caballero2016}%
  \BibitemOpen
  \bibfield  {author} {\bibinfo {author} {\bibfnamefont {A.}~\bibnamefont
  {{Marcos-Caballero}}}, \bibinfo {author} {\bibfnamefont {R.}~\bibnamefont
  {{Fern{\'a}ndez-Cobos}}}, \bibinfo {author} {\bibfnamefont {E.}~\bibnamefont
  {{Mart{\'\i}nez-Gonz{\'a}lez}}},\ and\ \bibinfo {author} {\bibfnamefont
  {P.}~\bibnamefont {{Vielva}}},\ }\bibfield  {title} {\bibinfo {title} {{On
  the void explanation of the Cold Spot}},\ }\href
  {https://doi.org/10.1093/mnrasl/slw063} {\bibfield  {journal} {\bibinfo
  {journal} {\mnras}\ }\textbf {\bibinfo {volume} {460}},\ \bibinfo {pages}
  {L15} (\bibinfo {year} {2016})},\ \Eprint {https://arxiv.org/abs/1510.09076}
  {arXiv:1510.09076 [astro-ph.CO]} \BibitemShut {NoStop}%
\bibitem [{\citenamefont {{Naidoo}}\ \emph {et~al.}(2017)\citenamefont
  {{Naidoo}}, \citenamefont {{Benoit-L{\'e}vy}},\ and\ \citenamefont
  {{Lahav}}}]{Naidoo2017}%
  \BibitemOpen
  \bibfield  {author} {\bibinfo {author} {\bibfnamefont {K.}~\bibnamefont
  {{Naidoo}}}, \bibinfo {author} {\bibfnamefont {A.}~\bibnamefont
  {{Benoit-L{\'e}vy}}},\ and\ \bibinfo {author} {\bibfnamefont
  {O.}~\bibnamefont {{Lahav}}},\ }\bibfield  {title} {\bibinfo {title} {{The
  cosmic microwave background Cold Spot anomaly: the impact of sky masking and
  the expected contribution from the integrated Sachs-Wolfe effect}},\ }\href
  {https://doi.org/10.1093/mnrasl/slx140} {\bibfield  {journal} {\bibinfo
  {journal} {\mnras}\ }\textbf {\bibinfo {volume} {472}},\ \bibinfo {pages}
  {L65} (\bibinfo {year} {2017})},\ \Eprint {https://arxiv.org/abs/1703.07894}
  {arXiv:1703.07894 [astro-ph.CO]} \BibitemShut {NoStop}%
\bibitem [{\citenamefont {{Kov{\'a}cs}}\ \emph {et~al.}(2020)\citenamefont
  {{Kov{\'a}cs}}, \citenamefont {{Beck}}, \citenamefont {{Szapudi}},
  \citenamefont {{Csabai}}, \citenamefont {{R{\'a}cz}},\ and\ \citenamefont
  {{Dobos}}}]{Kovacs2020}%
  \BibitemOpen
  \bibfield  {author} {\bibinfo {author} {\bibfnamefont {A.}~\bibnamefont
  {{Kov{\'a}cs}}}, \bibinfo {author} {\bibfnamefont {R.}~\bibnamefont
  {{Beck}}}, \bibinfo {author} {\bibfnamefont {I.}~\bibnamefont {{Szapudi}}},
  \bibinfo {author} {\bibfnamefont {I.}~\bibnamefont {{Csabai}}}, \bibinfo
  {author} {\bibfnamefont {G.}~\bibnamefont {{R{\'a}cz}}},\ and\ \bibinfo
  {author} {\bibfnamefont {L.}~\bibnamefont {{Dobos}}},\ }\bibfield  {title}
  {\bibinfo {title} {{A common explanation of the Hubble tension and anomalous
  cold spots in the CMB}},\ }\href {https://doi.org/10.1093/mnras/staa2631}
  {\bibfield  {journal} {\bibinfo  {journal} {\mnras}\ }\textbf {\bibinfo
  {volume} {499}},\ \bibinfo {pages} {320} (\bibinfo {year} {2020})},\ \Eprint
  {https://arxiv.org/abs/2004.02937} {arXiv:2004.02937 [astro-ph.CO]}
  \BibitemShut {NoStop}%
\bibitem [{\citenamefont {{Kov{\'a}cs}}\ \emph
  {et~al.}(2022{\natexlab{b}})\citenamefont {{Kov{\'a}cs}}, \citenamefont
  {{Jeffrey}}, \citenamefont {{Gatti}}, \citenamefont {{Chang}}, \citenamefont
  {{Whiteway}}, \citenamefont {{Hamaus}}, \citenamefont {{Lahav}},
  \citenamefont {{Pollina}}, \citenamefont {{Bacon}}, \citenamefont
  {{Kacprzak}} \emph {et~al.}}]{KovacsCS2022}%
  \BibitemOpen
  \bibfield  {author} {\bibinfo {author} {\bibfnamefont {A.}~\bibnamefont
  {{Kov{\'a}cs}}}, \bibinfo {author} {\bibfnamefont {N.}~\bibnamefont
  {{Jeffrey}}}, \bibinfo {author} {\bibfnamefont {M.}~\bibnamefont {{Gatti}}},
  \bibinfo {author} {\bibfnamefont {C.}~\bibnamefont {{Chang}}}, \bibinfo
  {author} {\bibfnamefont {L.}~\bibnamefont {{Whiteway}}}, \bibinfo {author}
  {\bibfnamefont {N.}~\bibnamefont {{Hamaus}}}, \bibinfo {author}
  {\bibfnamefont {O.}~\bibnamefont {{Lahav}}}, \bibinfo {author} {\bibfnamefont
  {G.}~\bibnamefont {{Pollina}}}, \bibinfo {author} {\bibfnamefont
  {D.}~\bibnamefont {{Bacon}}}, \bibinfo {author} {\bibfnamefont
  {T.}~\bibnamefont {{Kacprzak}}}, \emph {et~al.},\ }\bibfield  {title}
  {\bibinfo {title} {{The DES view of the Eridanus supervoid and the CMB cold
  spot}},\ }\href {https://doi.org/10.1093/mnras/stab3309} {\bibfield
  {journal} {\bibinfo  {journal} {\mnras}\ }\textbf {\bibinfo {volume} {510}},\
  \bibinfo {pages} {216} (\bibinfo {year} {2022}{\natexlab{b}})},\ \Eprint
  {https://arxiv.org/abs/2112.07699} {arXiv:2112.07699 [astro-ph.CO]}
  \BibitemShut {NoStop}%
\bibitem [{\citenamefont {{Torrado}}\ and\ \citenamefont
  {{Lewis}}(2019)}]{CobayaSoftware2019}%
  \BibitemOpen
  \bibfield  {author} {\bibinfo {author} {\bibfnamefont {J.}~\bibnamefont
  {{Torrado}}}\ and\ \bibinfo {author} {\bibfnamefont {A.}~\bibnamefont
  {{Lewis}}},\ }\href@noop {} {\bibinfo {title} {{Cobaya: Bayesian analysis in
  cosmology}}},\ \bibinfo {howpublished} {Astrophysics Source Code Library,
  record ascl:1910.019} (\bibinfo {year} {2019}),\ \Eprint
  {https://arxiv.org/abs/1910.019} {ascl:1910.019} \BibitemShut {NoStop}%
\bibitem [{\citenamefont {{Torrado}}\ and\ \citenamefont
  {{Lewis}}(2021)}]{CobayaPaper2021}%
  \BibitemOpen
  \bibfield  {author} {\bibinfo {author} {\bibfnamefont {J.}~\bibnamefont
  {{Torrado}}}\ and\ \bibinfo {author} {\bibfnamefont {A.}~\bibnamefont
  {{Lewis}}},\ }\bibfield  {title} {\bibinfo {title} {{Cobaya: code for
  Bayesian analysis of hierarchical physical models}},\ }\href
  {https://doi.org/10.1088/1475-7516/2021/05/057} {\bibfield  {journal}
  {\bibinfo  {journal} {\jcap}\ }\textbf {\bibinfo {volume} {2021}},\ \bibinfo
  {eid} {057} (\bibinfo {year} {2021})},\ \Eprint
  {https://arxiv.org/abs/2005.05290} {arXiv:2005.05290 [astro-ph.IM]}
  \BibitemShut {NoStop}%
\bibitem [{\citenamefont {{Planck Collaboration}}\ \emph
  {et~al.}(2020{\natexlab{b}})\citenamefont {{Planck Collaboration}},
  \citenamefont {{Aghanim}}, \citenamefont {{Akrami}}, \citenamefont
  {{Ashdown}}, \citenamefont {{Aumont}}, \citenamefont {{Baccigalupi}},
  \citenamefont {{Ballardini}}, \citenamefont {{Banday}}, \citenamefont
  {{Barreiro}}, \citenamefont {{Bartolo}}, \citenamefont {{Basak}},
  \citenamefont {{Benabed}}, \citenamefont {{Bernard}}, \citenamefont
  {{Bersanelli}}, \citenamefont {{Bielewicz}} \emph {et~al.}}]{Aghanim2019}%
  \BibitemOpen
  \bibfield  {author} {\bibinfo {author} {\bibnamefont {{Planck
  Collaboration}}}, \bibinfo {author} {\bibfnamefont {N.}~\bibnamefont
  {{Aghanim}}}, \bibinfo {author} {\bibfnamefont {Y.}~\bibnamefont {{Akrami}}},
  \bibinfo {author} {\bibfnamefont {M.}~\bibnamefont {{Ashdown}}}, \bibinfo
  {author} {\bibfnamefont {J.}~\bibnamefont {{Aumont}}}, \bibinfo {author}
  {\bibfnamefont {C.}~\bibnamefont {{Baccigalupi}}}, \bibinfo {author}
  {\bibfnamefont {M.}~\bibnamefont {{Ballardini}}}, \bibinfo {author}
  {\bibfnamefont {A.~J.}\ \bibnamefont {{Banday}}}, \bibinfo {author}
  {\bibfnamefont {R.~B.}\ \bibnamefont {{Barreiro}}}, \bibinfo {author}
  {\bibfnamefont {N.}~\bibnamefont {{Bartolo}}}, \bibinfo {author}
  {\bibfnamefont {S.}~\bibnamefont {{Basak}}}, \bibinfo {author} {\bibfnamefont
  {K.}~\bibnamefont {{Benabed}}}, \bibinfo {author} {\bibfnamefont {J.~P.}\
  \bibnamefont {{Bernard}}}, \bibinfo {author} {\bibfnamefont {M.}~\bibnamefont
  {{Bersanelli}}}, \bibinfo {author} {\bibfnamefont {P.}~\bibnamefont
  {{Bielewicz}}}, \emph {et~al.},\ }\bibfield  {title} {\bibinfo {title}
  {{Planck 2018 results. V. CMB power spectra and likelihoods}},\ }\href
  {https://doi.org/10.1051/0004-6361/201936386} {\bibfield  {journal} {\bibinfo
   {journal} {\aap}\ }\textbf {\bibinfo {volume} {641}},\ \bibinfo {eid} {A5}
  (\bibinfo {year} {2020}{\natexlab{b}})},\ \Eprint
  {https://arxiv.org/abs/1907.12875} {arXiv:1907.12875 [astro-ph.CO]}
  \BibitemShut {NoStop}%
\bibitem [{\citenamefont {{Scolnic}}\ \emph {et~al.}(2018)\citenamefont
  {{Scolnic}}, \citenamefont {{Jones}}, \citenamefont {{Rest}}, \citenamefont
  {{Pan}}, \citenamefont {{Chornock}}, \citenamefont {{Foley}}, \citenamefont
  {{Huber}}, \citenamefont {{Kessler}}, \citenamefont {{Narayan}},
  \citenamefont {{Riess}}, \citenamefont {{Rodney}}, \citenamefont {{Berger}},
  \citenamefont {{Brout}}, \citenamefont {{Challis}}, \citenamefont {{Drout}}
  \emph {et~al.}}]{Scolnic2017}%
  \BibitemOpen
  \bibfield  {author} {\bibinfo {author} {\bibfnamefont {D.~M.}\ \bibnamefont
  {{Scolnic}}}, \bibinfo {author} {\bibfnamefont {D.~O.}\ \bibnamefont
  {{Jones}}}, \bibinfo {author} {\bibfnamefont {A.}~\bibnamefont {{Rest}}},
  \bibinfo {author} {\bibfnamefont {Y.~C.}\ \bibnamefont {{Pan}}}, \bibinfo
  {author} {\bibfnamefont {R.}~\bibnamefont {{Chornock}}}, \bibinfo {author}
  {\bibfnamefont {R.~J.}\ \bibnamefont {{Foley}}}, \bibinfo {author}
  {\bibfnamefont {M.~E.}\ \bibnamefont {{Huber}}}, \bibinfo {author}
  {\bibfnamefont {R.}~\bibnamefont {{Kessler}}}, \bibinfo {author}
  {\bibfnamefont {G.}~\bibnamefont {{Narayan}}}, \bibinfo {author}
  {\bibfnamefont {A.~G.}\ \bibnamefont {{Riess}}}, \bibinfo {author}
  {\bibfnamefont {S.}~\bibnamefont {{Rodney}}}, \bibinfo {author}
  {\bibfnamefont {E.}~\bibnamefont {{Berger}}}, \bibinfo {author}
  {\bibfnamefont {D.~J.}\ \bibnamefont {{Brout}}}, \bibinfo {author}
  {\bibfnamefont {P.~J.}\ \bibnamefont {{Challis}}}, \bibinfo {author}
  {\bibfnamefont {M.}~\bibnamefont {{Drout}}}, \emph {et~al.},\ }\bibfield
  {title} {\bibinfo {title} {{The Complete Light-curve Sample of
  Spectroscopically Confirmed SNe Ia from Pan-STARRS1 and Cosmological
  Constraints from the Combined Pantheon Sample}},\ }\href
  {https://doi.org/10.3847/1538-4357/aab9bb} {\bibfield  {journal} {\bibinfo
  {journal} {\apj}\ }\textbf {\bibinfo {volume} {859}},\ \bibinfo {eid} {101}
  (\bibinfo {year} {2018})},\ \Eprint {https://arxiv.org/abs/1710.00845}
  {arXiv:1710.00845 [astro-ph.CO]} \BibitemShut {NoStop}%
\bibitem [{\citenamefont {{Riess}}\ \emph
  {et~al.}(2021{\natexlab{b}})\citenamefont {{Riess}}, \citenamefont
  {{Casertano}}, \citenamefont {{Yuan}}, \citenamefont {{Bowers}},
  \citenamefont {{Macri}}, \citenamefont {{Zinn}},\ and\ \citenamefont
  {{Scolnic}}}]{RiessMb2021}%
  \BibitemOpen
  \bibfield  {author} {\bibinfo {author} {\bibfnamefont {A.~G.}\ \bibnamefont
  {{Riess}}}, \bibinfo {author} {\bibfnamefont {S.}~\bibnamefont
  {{Casertano}}}, \bibinfo {author} {\bibfnamefont {W.}~\bibnamefont {{Yuan}}},
  \bibinfo {author} {\bibfnamefont {J.~B.}\ \bibnamefont {{Bowers}}}, \bibinfo
  {author} {\bibfnamefont {L.}~\bibnamefont {{Macri}}}, \bibinfo {author}
  {\bibfnamefont {J.~C.}\ \bibnamefont {{Zinn}}},\ and\ \bibinfo {author}
  {\bibfnamefont {D.}~\bibnamefont {{Scolnic}}},\ }\bibfield  {title} {\bibinfo
  {title} {{Cosmic Distances Calibrated to 1\% Precision with Gaia EDR3
  Parallaxes and Hubble Space Telescope Photometry of 75 Milky Way Cepheids
  Confirm Tension with {\ensuremath{\Lambda}}CDM}},\ }\href
  {https://doi.org/10.3847/2041-8213/abdbaf} {\bibfield  {journal} {\bibinfo
  {journal} {\apjl}\ }\textbf {\bibinfo {volume} {908}},\ \bibinfo {eid} {L6}
  (\bibinfo {year} {2021}{\natexlab{b}})},\ \Eprint
  {https://arxiv.org/abs/2012.08534} {arXiv:2012.08534 [astro-ph.CO]}
  \BibitemShut {NoStop}%
\bibitem [{\citenamefont {{Beutler}}\ \emph {et~al.}(2012)\citenamefont
  {{Beutler}}, \citenamefont {{Blake}}, \citenamefont {{Colless}},
  \citenamefont {{Jones}}, \citenamefont {{Staveley-Smith}}, \citenamefont
  {{Poole}}, \citenamefont {{Campbell}}, \citenamefont {{Parker}},
  \citenamefont {{Saunders}},\ and\ \citenamefont {{Watson}}}]{Beutler2012}%
  \BibitemOpen
  \bibfield  {author} {\bibinfo {author} {\bibfnamefont {F.}~\bibnamefont
  {{Beutler}}}, \bibinfo {author} {\bibfnamefont {C.}~\bibnamefont {{Blake}}},
  \bibinfo {author} {\bibfnamefont {M.}~\bibnamefont {{Colless}}}, \bibinfo
  {author} {\bibfnamefont {D.~H.}\ \bibnamefont {{Jones}}}, \bibinfo {author}
  {\bibfnamefont {L.}~\bibnamefont {{Staveley-Smith}}}, \bibinfo {author}
  {\bibfnamefont {G.~B.}\ \bibnamefont {{Poole}}}, \bibinfo {author}
  {\bibfnamefont {L.}~\bibnamefont {{Campbell}}}, \bibinfo {author}
  {\bibfnamefont {Q.}~\bibnamefont {{Parker}}}, \bibinfo {author}
  {\bibfnamefont {W.}~\bibnamefont {{Saunders}}},\ and\ \bibinfo {author}
  {\bibfnamefont {F.}~\bibnamefont {{Watson}}},\ }\bibfield  {title} {\bibinfo
  {title} {{The 6dF Galaxy Survey: z{\ensuremath{\approx}} 0 measurements of
  the growth rate and {\ensuremath{\sigma}}$_{8}$}},\ }\href
  {https://doi.org/10.1111/j.1365-2966.2012.21136.x} {\bibfield  {journal}
  {\bibinfo  {journal} {\mnras}\ }\textbf {\bibinfo {volume} {423}},\ \bibinfo
  {pages} {3430} (\bibinfo {year} {2012})},\ \Eprint
  {https://arxiv.org/abs/1204.4725} {arXiv:1204.4725 [astro-ph.CO]}
  \BibitemShut {NoStop}%
\bibitem [{\citenamefont {Ross}\ \emph {et~al.}(2015)\citenamefont {Ross},
  \citenamefont {Samushia}, \citenamefont {Howlett}, \citenamefont {Percival},
  \citenamefont {Burden},\ and\ \citenamefont {Manera}}]{Ross2014}%
  \BibitemOpen
  \bibfield  {author} {\bibinfo {author} {\bibfnamefont {A.~J.}\ \bibnamefont
  {Ross}}, \bibinfo {author} {\bibfnamefont {L.}~\bibnamefont {Samushia}},
  \bibinfo {author} {\bibfnamefont {C.}~\bibnamefont {Howlett}}, \bibinfo
  {author} {\bibfnamefont {W.~J.}\ \bibnamefont {Percival}}, \bibinfo {author}
  {\bibfnamefont {A.}~\bibnamefont {Burden}},\ and\ \bibinfo {author}
  {\bibfnamefont {M.}~\bibnamefont {Manera}},\ }\bibfield  {title} {\bibinfo
  {title} {{The clustering of the SDSS DR7 main Galaxy sample – I. A 4 per
  cent distance measure at $z = 0.15$}},\ }\href
  {https://doi.org/10.1093/mnras/stv154} {\bibfield  {journal} {\bibinfo
  {journal} {Mon. Not. Roy. Astron. Soc.}\ }\textbf {\bibinfo {volume} {449}},\
  \bibinfo {pages} {835} (\bibinfo {year} {2015})},\ \Eprint
  {https://arxiv.org/abs/1409.3242} {arXiv:1409.3242 [astro-ph.CO]}
  \BibitemShut {NoStop}%
\bibitem [{\citenamefont {{Alam}}\ \emph {et~al.}(2017)\citenamefont {{Alam}},
  \citenamefont {{Ata}}, \citenamefont {{Bailey}}, \citenamefont {{Beutler}},
  \citenamefont {{Bizyaev}}, \citenamefont {{Blazek}}, \citenamefont
  {{Bolton}}, \citenamefont {{Brownstein}}, \citenamefont {{Burden}},
  \citenamefont {{Chuang}}, \citenamefont {{Comparat}}, \citenamefont
  {{Cuesta}}, \citenamefont {{Dawson}}, \citenamefont {{Eisenstein}},
  \citenamefont {{Escoffier}} \emph {et~al.}}]{Alam2016}%
  \BibitemOpen
  \bibfield  {author} {\bibinfo {author} {\bibfnamefont {S.}~\bibnamefont
  {{Alam}}}, \bibinfo {author} {\bibfnamefont {M.}~\bibnamefont {{Ata}}},
  \bibinfo {author} {\bibfnamefont {S.}~\bibnamefont {{Bailey}}}, \bibinfo
  {author} {\bibfnamefont {F.}~\bibnamefont {{Beutler}}}, \bibinfo {author}
  {\bibfnamefont {D.}~\bibnamefont {{Bizyaev}}}, \bibinfo {author}
  {\bibfnamefont {J.~A.}\ \bibnamefont {{Blazek}}}, \bibinfo {author}
  {\bibfnamefont {A.~S.}\ \bibnamefont {{Bolton}}}, \bibinfo {author}
  {\bibfnamefont {J.~R.}\ \bibnamefont {{Brownstein}}}, \bibinfo {author}
  {\bibfnamefont {A.}~\bibnamefont {{Burden}}}, \bibinfo {author}
  {\bibfnamefont {C.-H.}\ \bibnamefont {{Chuang}}}, \bibinfo {author}
  {\bibfnamefont {J.}~\bibnamefont {{Comparat}}}, \bibinfo {author}
  {\bibfnamefont {A.~J.}\ \bibnamefont {{Cuesta}}}, \bibinfo {author}
  {\bibfnamefont {K.~S.}\ \bibnamefont {{Dawson}}}, \bibinfo {author}
  {\bibfnamefont {D.~J.}\ \bibnamefont {{Eisenstein}}}, \bibinfo {author}
  {\bibfnamefont {S.}~\bibnamefont {{Escoffier}}}, \emph {et~al.},\ }\bibfield
  {title} {\bibinfo {title} {{The clustering of galaxies in the completed
  SDSS-III Baryon Oscillation Spectroscopic Survey: cosmological analysis of
  the DR12 galaxy sample}},\ }\href {https://doi.org/10.1093/mnras/stx721}
  {\bibfield  {journal} {\bibinfo  {journal} {\mnras}\ }\textbf {\bibinfo
  {volume} {470}},\ \bibinfo {pages} {2617} (\bibinfo {year} {2017})},\ \Eprint
  {https://arxiv.org/abs/1607.03155} {arXiv:1607.03155 [astro-ph.CO]}
  \BibitemShut {NoStop}%
\bibitem [{\citenamefont {Lewis}(2013)}]{Lewis2013}%
  \BibitemOpen
  \bibfield  {author} {\bibinfo {author} {\bibfnamefont {A.}~\bibnamefont
  {Lewis}},\ }\bibfield  {title} {\bibinfo {title} {{Efficient sampling of fast
  and slow cosmological parameters}},\ }\href
  {https://doi.org/10.1103/PhysRevD.87.103529} {\bibfield  {journal} {\bibinfo
  {journal} {Phys. Rev.}\ }\textbf {\bibinfo {volume} {D87}},\ \bibinfo {pages}
  {103529} (\bibinfo {year} {2013})},\ \Eprint
  {https://arxiv.org/abs/1304.4473} {arXiv:1304.4473 [astro-ph.CO]}
  \BibitemShut {NoStop}%
\bibitem [{\citenamefont {Akaike}(1974)}]{Akaike1974}%
  \BibitemOpen
  \bibfield  {author} {\bibinfo {author} {\bibfnamefont {H.}~\bibnamefont
  {Akaike}},\ }\bibfield  {title} {\bibinfo {title} {A new look at the
  statistical model identification},\ }\href@noop {} {\bibfield  {journal}
  {\bibinfo  {journal} {IEEE transactions on automatic control}\ }\textbf
  {\bibinfo {volume} {19}},\ \bibinfo {pages} {716} (\bibinfo {year}
  {1974})}\BibitemShut {NoStop}%
\bibitem [{\citenamefont {{Heymans}}\ \emph {et~al.}(2021)\citenamefont
  {{Heymans}}, \citenamefont {{Tr{\"o}ster}}, \citenamefont {{Asgari}},
  \citenamefont {{Blake}}, \citenamefont {{Hildebrandt}}, \citenamefont
  {{Joachimi}}, \citenamefont {{Kuijken}}, \citenamefont {{Lin}}, \citenamefont
  {{S{\'a}nchez}}, \citenamefont {{van den Busch}}, \citenamefont {{Wright}},
  \citenamefont {{Amon}}, \citenamefont {{Bilicki}}, \citenamefont {{de Jong}},
  \citenamefont {{Crocce}} \emph {et~al.}}]{Kids2021}%
  \BibitemOpen
  \bibfield  {author} {\bibinfo {author} {\bibfnamefont {C.}~\bibnamefont
  {{Heymans}}}, \bibinfo {author} {\bibfnamefont {T.}~\bibnamefont
  {{Tr{\"o}ster}}}, \bibinfo {author} {\bibfnamefont {M.}~\bibnamefont
  {{Asgari}}}, \bibinfo {author} {\bibfnamefont {C.}~\bibnamefont {{Blake}}},
  \bibinfo {author} {\bibfnamefont {H.}~\bibnamefont {{Hildebrandt}}}, \bibinfo
  {author} {\bibfnamefont {B.}~\bibnamefont {{Joachimi}}}, \bibinfo {author}
  {\bibfnamefont {K.}~\bibnamefont {{Kuijken}}}, \bibinfo {author}
  {\bibfnamefont {C.-A.}\ \bibnamefont {{Lin}}}, \bibinfo {author}
  {\bibfnamefont {A.~G.}\ \bibnamefont {{S{\'a}nchez}}}, \bibinfo {author}
  {\bibfnamefont {J.~L.}\ \bibnamefont {{van den Busch}}}, \bibinfo {author}
  {\bibfnamefont {A.~H.}\ \bibnamefont {{Wright}}}, \bibinfo {author}
  {\bibfnamefont {A.}~\bibnamefont {{Amon}}}, \bibinfo {author} {\bibfnamefont
  {M.}~\bibnamefont {{Bilicki}}}, \bibinfo {author} {\bibfnamefont
  {J.}~\bibnamefont {{de Jong}}}, \bibinfo {author} {\bibfnamefont
  {M.}~\bibnamefont {{Crocce}}}, \emph {et~al.},\ }\bibfield  {title} {\bibinfo
  {title} {{KiDS-1000 Cosmology: Multi-probe weak gravitational lensing and
  spectroscopic galaxy clustering constraints}},\ }\href
  {https://doi.org/10.1051/0004-6361/202039063} {\bibfield  {journal} {\bibinfo
   {journal} {\aap}\ }\textbf {\bibinfo {volume} {646}},\ \bibinfo {eid} {A140}
  (\bibinfo {year} {2021})},\ \Eprint {https://arxiv.org/abs/2007.15632}
  {arXiv:2007.15632 [astro-ph.CO]} \BibitemShut {NoStop}%
\bibitem [{\citenamefont {{Planck Collaboration}}\ \emph
  {et~al.}(2020{\natexlab{c}})\citenamefont {{Planck Collaboration}},
  \citenamefont {{Aghanim}}, \citenamefont {{Akrami}}, \citenamefont
  {{Ashdown}}, \citenamefont {{Aumont}}, \citenamefont {{Baccigalupi}},
  \citenamefont {{Ballardini}}, \citenamefont {{Banday}}, \citenamefont
  {{Barreiro}}, \citenamefont {{Bartolo}}, \citenamefont {{Basak}},
  \citenamefont {{Benabed}}, \citenamefont {{Bernard}}, \citenamefont
  {{Bersanelli}}, \citenamefont {{Bielewicz}} \emph {et~al.}}]{Aghanim2018}%
  \BibitemOpen
  \bibfield  {author} {\bibinfo {author} {\bibnamefont {{Planck
  Collaboration}}}, \bibinfo {author} {\bibfnamefont {N.}~\bibnamefont
  {{Aghanim}}}, \bibinfo {author} {\bibfnamefont {Y.}~\bibnamefont {{Akrami}}},
  \bibinfo {author} {\bibfnamefont {M.}~\bibnamefont {{Ashdown}}}, \bibinfo
  {author} {\bibfnamefont {J.}~\bibnamefont {{Aumont}}}, \bibinfo {author}
  {\bibfnamefont {C.}~\bibnamefont {{Baccigalupi}}}, \bibinfo {author}
  {\bibfnamefont {M.}~\bibnamefont {{Ballardini}}}, \bibinfo {author}
  {\bibfnamefont {A.~J.}\ \bibnamefont {{Banday}}}, \bibinfo {author}
  {\bibfnamefont {R.~B.}\ \bibnamefont {{Barreiro}}}, \bibinfo {author}
  {\bibfnamefont {N.}~\bibnamefont {{Bartolo}}}, \bibinfo {author}
  {\bibfnamefont {S.}~\bibnamefont {{Basak}}}, \bibinfo {author} {\bibfnamefont
  {K.}~\bibnamefont {{Benabed}}}, \bibinfo {author} {\bibfnamefont {J.~P.}\
  \bibnamefont {{Bernard}}}, \bibinfo {author} {\bibfnamefont {M.}~\bibnamefont
  {{Bersanelli}}}, \bibinfo {author} {\bibfnamefont {P.}~\bibnamefont
  {{Bielewicz}}}, \emph {et~al.},\ }\bibfield  {title} {\bibinfo {title}
  {{Planck 2018 results. VIII. Gravitational lensing}},\ }\href
  {https://doi.org/10.1051/0004-6361/201833886} {\bibfield  {journal} {\bibinfo
   {journal} {\aap}\ }\textbf {\bibinfo {volume} {641}},\ \bibinfo {eid} {A8}
  (\bibinfo {year} {2020}{\natexlab{c}})},\ \Eprint
  {https://arxiv.org/abs/1807.06210} {arXiv:1807.06210 [astro-ph.CO]}
  \BibitemShut {NoStop}%
\bibitem [{\citenamefont {{Efstathiou}}(2021)}]{Efstathiou2021}%
  \BibitemOpen
  \bibfield  {author} {\bibinfo {author} {\bibfnamefont {G.}~\bibnamefont
  {{Efstathiou}}},\ }\bibfield  {title} {\bibinfo {title} {{To H$_{0}$ or not
  to H$_{0}$?}},\ }\href {https://doi.org/10.1093/mnras/stab1588} {\bibfield
  {journal} {\bibinfo  {journal} {\mnras}\ }\textbf {\bibinfo {volume} {505}},\
  \bibinfo {pages} {3866} (\bibinfo {year} {2021})},\ \Eprint
  {https://arxiv.org/abs/2103.08723} {arXiv:2103.08723 [astro-ph.CO]}
  \BibitemShut {NoStop}%
\bibitem [{\citenamefont {{Camarena}}\ and\ \citenamefont
  {{Marra}}(2021)}]{Camarena2021}%
  \BibitemOpen
  \bibfield  {author} {\bibinfo {author} {\bibfnamefont {D.}~\bibnamefont
  {{Camarena}}}\ and\ \bibinfo {author} {\bibfnamefont {V.}~\bibnamefont
  {{Marra}}},\ }\bibfield  {title} {\bibinfo {title} {{On the use of the local
  prior on the absolute magnitude of Type Ia supernovae in cosmological
  inference}},\ }\href {https://doi.org/10.1093/mnras/stab1200} {\bibfield
  {journal} {\bibinfo  {journal} {\mnras}\ }\textbf {\bibinfo {volume} {504}},\
  \bibinfo {pages} {5164} (\bibinfo {year} {2021})},\ \Eprint
  {https://arxiv.org/abs/2101.08641} {arXiv:2101.08641 [astro-ph.CO]}
  \BibitemShut {NoStop}%
\bibitem [{\citenamefont {{Poulin}}\ \emph {et~al.}(2019)\citenamefont
  {{Poulin}}, \citenamefont {{Smith}}, \citenamefont {{Karwal}},\ and\
  \citenamefont {{Kamionkowski}}}]{EDE}%
  \BibitemOpen
  \bibfield  {author} {\bibinfo {author} {\bibfnamefont {V.}~\bibnamefont
  {{Poulin}}}, \bibinfo {author} {\bibfnamefont {T.~L.}\ \bibnamefont
  {{Smith}}}, \bibinfo {author} {\bibfnamefont {T.}~\bibnamefont {{Karwal}}},\
  and\ \bibinfo {author} {\bibfnamefont {M.}~\bibnamefont {{Kamionkowski}}},\
  }\bibfield  {title} {\bibinfo {title} {{Early Dark Energy can Resolve the
  Hubble Tension}},\ }\href {https://doi.org/10.1103/PhysRevLett.122.221301}
  {\bibfield  {journal} {\bibinfo  {journal} {\prl}\ }\textbf {\bibinfo
  {volume} {122}},\ \bibinfo {eid} {221301} (\bibinfo {year} {2019})},\ \Eprint
  {https://arxiv.org/abs/1811.04083} {arXiv:1811.04083 [astro-ph.CO]}
  \BibitemShut {NoStop}%
\bibitem [{\citenamefont {{Vagnozzi}}(2021)}]{Vagnozzi2021}%
  \BibitemOpen
  \bibfield  {author} {\bibinfo {author} {\bibfnamefont {S.}~\bibnamefont
  {{Vagnozzi}}},\ }\bibfield  {title} {\bibinfo {title} {{Consistency tests of
  {\ensuremath{\Lambda}} CDM from the early integrated Sachs-Wolfe effect:
  Implications for early-time new physics and the Hubble tension}},\ }\href
  {https://doi.org/10.1103/PhysRevD.104.063524} {\bibfield  {journal} {\bibinfo
   {journal} {\prd}\ }\textbf {\bibinfo {volume} {104}},\ \bibinfo {eid}
  {063524} (\bibinfo {year} {2021})},\ \Eprint
  {https://arxiv.org/abs/2105.10425} {arXiv:2105.10425 [astro-ph.CO]}
  \BibitemShut {NoStop}%
\bibitem [{\citenamefont {{Garc{\'\i}a Escudero}}\ \emph
  {et~al.}(2022)\citenamefont {{Garc{\'\i}a Escudero}}, \citenamefont {{Kuo}},
  \citenamefont {{Keeley}},\ and\ \citenamefont {{Abazajian}}}]{Escudero2022}%
  \BibitemOpen
  \bibfield  {author} {\bibinfo {author} {\bibfnamefont {H.}~\bibnamefont
  {{Garc{\'\i}a Escudero}}}, \bibinfo {author} {\bibfnamefont {J.-L.}\
  \bibnamefont {{Kuo}}}, \bibinfo {author} {\bibfnamefont {R.~E.}\ \bibnamefont
  {{Keeley}}},\ and\ \bibinfo {author} {\bibfnamefont {K.~N.}\ \bibnamefont
  {{Abazajian}}},\ }\bibfield  {title} {\bibinfo {title} {{Early versus Phantom
  Dark Energy, Self-Interacting, Extra, or Massive Neutrinos, Primordial
  Magnetic Fields, or a Curved Universe: An Exploration of Possible Solutions
  to the $H_0$ and $\sigma_8$ Problems}},\ }\href@noop {} {\bibfield  {journal}
  {\bibinfo  {journal} {arXiv e-prints}\ ,\ \bibinfo {eid} {arXiv:2208.14435}}
  (\bibinfo {year} {2022})},\ \Eprint {https://arxiv.org/abs/2208.14435}
  {arXiv:2208.14435 [astro-ph.CO]} \BibitemShut {NoStop}%
\bibitem [{\citenamefont {{Pourtsidou}}\ and\ \citenamefont
  {{Tram}}(2016)}]{Pourtsidou2016}%
  \BibitemOpen
  \bibfield  {author} {\bibinfo {author} {\bibfnamefont {A.}~\bibnamefont
  {{Pourtsidou}}}\ and\ \bibinfo {author} {\bibfnamefont {T.}~\bibnamefont
  {{Tram}}},\ }\bibfield  {title} {\bibinfo {title} {{Reconciling CMB and
  structure growth measurements with dark energy interactions}},\ }\href
  {https://doi.org/10.1103/PhysRevD.94.043518} {\bibfield  {journal} {\bibinfo
  {journal} {\prd}\ }\textbf {\bibinfo {volume} {94}},\ \bibinfo {eid} {043518}
  (\bibinfo {year} {2016})},\ \Eprint {https://arxiv.org/abs/1604.04222}
  {arXiv:1604.04222 [astro-ph.CO]} \BibitemShut {NoStop}%
\bibitem [{\citenamefont {{Kumar}}\ and\ \citenamefont
  {{Nunes}}(2016)}]{Kumar2016}%
  \BibitemOpen
  \bibfield  {author} {\bibinfo {author} {\bibfnamefont {S.}~\bibnamefont
  {{Kumar}}}\ and\ \bibinfo {author} {\bibfnamefont {R.~C.}\ \bibnamefont
  {{Nunes}}},\ }\bibfield  {title} {\bibinfo {title} {{Probing the interaction
  between dark matter and dark energy in the presence of massive neutrinos}},\
  }\href {https://doi.org/10.1103/PhysRevD.94.123511} {\bibfield  {journal}
  {\bibinfo  {journal} {\prd}\ }\textbf {\bibinfo {volume} {94}},\ \bibinfo
  {eid} {123511} (\bibinfo {year} {2016})},\ \Eprint
  {https://arxiv.org/abs/1608.02454} {arXiv:1608.02454 [astro-ph.CO]}
  \BibitemShut {NoStop}%
\bibitem [{\citenamefont {{Kumar}}\ \emph {et~al.}(2019)\citenamefont
  {{Kumar}}, \citenamefont {{Nunes}},\ and\ \citenamefont
  {{Yadav}}}]{Kumar2019}%
  \BibitemOpen
  \bibfield  {author} {\bibinfo {author} {\bibfnamefont {S.}~\bibnamefont
  {{Kumar}}}, \bibinfo {author} {\bibfnamefont {R.~C.}\ \bibnamefont
  {{Nunes}}},\ and\ \bibinfo {author} {\bibfnamefont {S.~K.}\ \bibnamefont
  {{Yadav}}},\ }\bibfield  {title} {\bibinfo {title} {{Dark sector interaction:
  a remedy of the tensions between CMB and LSS data}},\ }\href
  {https://doi.org/10.1140/epjc/s10052-019-7087-7} {\bibfield  {journal}
  {\bibinfo  {journal} {European Physical Journal C}\ }\textbf {\bibinfo
  {volume} {79}},\ \bibinfo {eid} {576} (\bibinfo {year} {2019})},\ \Eprint
  {https://arxiv.org/abs/1903.04865} {arXiv:1903.04865 [astro-ph.CO]}
  \BibitemShut {NoStop}%
\bibitem [{\citenamefont {{Di Valentino}}\ \emph {et~al.}(2020)\citenamefont
  {{Di Valentino}}, \citenamefont {{Melchiorri}}, \citenamefont {{Mena}},\ and\
  \citenamefont {{Vagnozzi}}}]{DiValentino2020}%
  \BibitemOpen
  \bibfield  {author} {\bibinfo {author} {\bibfnamefont {E.}~\bibnamefont {{Di
  Valentino}}}, \bibinfo {author} {\bibfnamefont {A.}~\bibnamefont
  {{Melchiorri}}}, \bibinfo {author} {\bibfnamefont {O.}~\bibnamefont
  {{Mena}}},\ and\ \bibinfo {author} {\bibfnamefont {S.}~\bibnamefont
  {{Vagnozzi}}},\ }\bibfield  {title} {\bibinfo {title} {{Interacting dark
  energy in the early 2020s: A promising solution to the H$_{0}$ and cosmic
  shear tensions}},\ }\href {https://doi.org/10.1016/j.dark.2020.100666}
  {\bibfield  {journal} {\bibinfo  {journal} {Physics of the Dark Universe}\
  }\textbf {\bibinfo {volume} {30}},\ \bibinfo {eid} {100666} (\bibinfo {year}
  {2020})},\ \Eprint {https://arxiv.org/abs/1908.04281} {arXiv:1908.04281
  [astro-ph.CO]} \BibitemShut {NoStop}%
\bibitem [{\citenamefont {{Lucca}}\ and\ \citenamefont
  {{Hooper}}(2020)}]{Lucca2020}%
  \BibitemOpen
  \bibfield  {author} {\bibinfo {author} {\bibfnamefont {M.}~\bibnamefont
  {{Lucca}}}\ and\ \bibinfo {author} {\bibfnamefont {D.~C.}\ \bibnamefont
  {{Hooper}}},\ }\bibfield  {title} {\bibinfo {title} {{Shedding light on dark
  matter-dark energy interactions}},\ }\href
  {https://doi.org/10.1103/PhysRevD.102.123502} {\bibfield  {journal} {\bibinfo
   {journal} {\prd}\ }\textbf {\bibinfo {volume} {102}},\ \bibinfo {eid}
  {123502} (\bibinfo {year} {2020})},\ \Eprint
  {https://arxiv.org/abs/2002.06127} {arXiv:2002.06127 [astro-ph.CO]}
  \BibitemShut {NoStop}%
\bibitem [{\citenamefont {{Poulin}}\ \emph {et~al.}(2022)\citenamefont
  {{Poulin}}, \citenamefont {{Bernal}}, \citenamefont {{Kovetz}},\ and\
  \citenamefont {{Kamionkowski}}}]{Poulin2022}%
  \BibitemOpen
  \bibfield  {author} {\bibinfo {author} {\bibfnamefont {V.}~\bibnamefont
  {{Poulin}}}, \bibinfo {author} {\bibfnamefont {J.~L.}\ \bibnamefont
  {{Bernal}}}, \bibinfo {author} {\bibfnamefont {E.}~\bibnamefont {{Kovetz}}},\
  and\ \bibinfo {author} {\bibfnamefont {M.}~\bibnamefont {{Kamionkowski}}},\
  }\bibfield  {title} {\bibinfo {title} {{The Sigma-8 Tension is a Drag}},\
  }\href@noop {} {\bibfield  {journal} {\bibinfo  {journal} {arXiv e-prints}\
  ,\ \bibinfo {eid} {arXiv:2209.06217}} (\bibinfo {year} {2022})},\ \Eprint
  {https://arxiv.org/abs/2209.06217} {arXiv:2209.06217 [astro-ph.CO]}
  \BibitemShut {NoStop}%
\bibitem [{\citenamefont {Lewis}\ and\ \citenamefont
  {Bridle}(2002)}]{Lewis2002}%
  \BibitemOpen
  \bibfield  {author} {\bibinfo {author} {\bibfnamefont {A.}~\bibnamefont
  {Lewis}}\ and\ \bibinfo {author} {\bibfnamefont {S.}~\bibnamefont {Bridle}},\
  }\bibfield  {title} {\bibinfo {title} {{Cosmological parameters from CMB and
  other data: A Monte Carlo approach}},\ }\href
  {https://doi.org/10.1103/PhysRevD.66.103511} {\bibfield  {journal} {\bibinfo
  {journal} {Phys. Rev.}\ }\textbf {\bibinfo {volume} {D66}},\ \bibinfo {pages}
  {103511} (\bibinfo {year} {2002})},\ \Eprint
  {https://arxiv.org/abs/astro-ph/0205436} {arXiv:astro-ph/0205436 [astro-ph]}
  \BibitemShut {NoStop}%
\bibitem [{\citenamefont {{Neal}}(2005)}]{Neal2005}%
  \BibitemOpen
  \bibfield  {author} {\bibinfo {author} {\bibfnamefont {R.~M.}\ \bibnamefont
  {{Neal}}},\ }\bibfield  {title} {\bibinfo {title} {{Taking Bigger Metropolis
  Steps by Dragging Fast Variables}},\ }\href
  {https://arxiv.org/abs/math/0502099} {\bibfield  {journal} {\bibinfo
  {journal} {ArXiv Mathematics e-prints}\ } (\bibinfo {year} {2005})},\ \Eprint
  {https://arxiv.org/abs/math/0502099} {math/0502099} \BibitemShut {NoStop}%
\end{thebibliography}%

\end{document}